\documentclass[12pt,a4paper]{article}
\usepackage[usenames,dvipsnames]{pstricks} 
\usepackage{afterpage}
\usepackage{epsfig}
\usepackage{pst-grad} 
\usepackage{pst-plot} 
\usepackage[a4paper]{geometry}
\usepackage{float}
\usepackage[stable]{footmisc}
\usepackage[utf8x]{inputenc}
\usepackage{a4wide}
\usepackage{amsmath,amssymb,amstext,amsthm,amssymb}
\usepackage{amsfonts}
\usepackage{framed}
\usepackage{graphicx}
\usepackage{color}
\usepackage{bbold}
\usepackage{bbm}
\usepackage{rotating} 
\usepackage{slashed} 
\usepackage{cancel} 
\usepackage{braket} 
\usepackage{amstext}
\usepackage{array}
\usepackage{setspace}
\usepackage{hyperref}
\usepackage{overpic}
\usepackage{bbm}
\usepackage[nosort]{cite}
\usepackage{lipsum}
\usepackage{dsfont}

\newcommand{\del}[0]{\partial}

\let\baraccent=\=
\renewcommand{\=}[1]{\stackrel{#1}{=}}

\DeclareSymbolFontAlphabet{\mathbb}{AMSb}
\begin{document}

\pagestyle{plain}
\pagenumbering{gobble} 

\makeatletter
\@addtoreset{equation}{section}
\makeatother
\renewcommand{\theequation}{\thesection.\arabic{equation}}
\pagestyle{empty}
\vspace{0.5cm}

\begin{center}
    
	{\LARGE \bf{$G_2$-manifolds from Diophantine equations} \\[4mm] }

\end{center}

\vspace{1cm}

\begin{center}
	\scalebox{0.95}[0.95]{{\fontsize{14}{30}\selectfont Jakob Moritz}}
\end{center}

\begin{center}
	\vspace{0.25 cm}
	
		\textsl{Department of Physics, University of Wisconsin-Madison, Madison, WI 53706, USA}\\

	\vspace{1cm}
	\normalsize{\bf Abstract} \\[8mm]

\end{center}

\begin{center}
	\begin{minipage}[h]{15.0cm}
		
		We argue that \emph{perturbatively flat vacua} (PFVs) introduced in \cite{Demirtas:2019sip} are dual to M-theory compactifications on $G_2$-manifolds, enabling the enumeration of potentially novel $G_2$-manifolds via solutions to Diophantine equations in type IIB flux quanta.
        Independently, we show that warping corrections to the effective action of type IIB flux vacua grow parametrically at large complex structure,
        and we demonstrate that these corrections can nonetheless be captured by a classical geometric computation in M-theory.
		
	\end{minipage}
\end{center}
\newpage
\setcounter{page}{1}
\pagestyle{plain}
\renewcommand{\thefootnote}{\arabic{footnote}}
\setcounter{footnote}{0}
%
%
\tableofcontents
\newpage
\pagenumbering{arabic} 
\setcounter{page}{1}

	\section{Introduction}\label{sec:Introduction}

    In order to understand quantum gravity in our universe, one can study compactifications of weakly coupled string theories to four dimensions. Many promising approaches have been developed over the past few decades, using both supersymmetric and non-supersymmetric critical string theories, as well as non-critical string theories.\footnote{See \cite{McAllister:2023vgy} and references therein.} A central goal in any approach is to make contact with cosmology, by identifying a four-dimensional metastable de Sitter vacuum with small vacuum energy.

    In this paper, we focus on the landscape of $4d$ $\mathcal{N}=1$ supergravity limits that arise from compactifications of type II string theories on Calabi-Yau orientifolds with background fluxes \cite{Giddings:2001yu,DeWolfe:2005uu}, and from compactifications of 11d M-theory on compact $G_2$-holonomy manifolds \cite{Dasgupta:1999ss,Beasley:2002db}. These compactifications typically involve many light moduli, and their (meta)stability can be studied via the effective F-term potential in the four-dimensional supergravity approximation.

    A particularly influential proposal for realizing metastable de Sitter vacua in this setting is the KKLT construction \cite{Kachru:2003aw}, which has been developed more explicitly in recent years \cite{Demirtas:2019sip,Demirtas:2020ffz,Alvarez-Garcia:2020pxd,Demirtas:2021nlu,McAllister:2024lnt}, while also attracting significant scrutiny \cite{Moritz:2017xto,Gautason:2018gln,Hamada:2018qef,Hamada:2019ack,Carta:2019rhx,Gautason:2019jwq,Bena:2019mte,Kachru:2019dvo,Randall:2019ent,Bena:2020xrh,Lust:2022lfc,Lust:2022xoq,Bena:2024are}. A concrete realization of this proposal was introduced in \cite{Demirtas:2019sip}: the starting point is a \emph{perturbatively flat vacuum} (PFV), defined as a compactification on a Calabi-Yau orientifold of O3/O7 type with three-form fluxes $F_3$ and $H_3$, such that the low-energy superpotential $W$ vanishes identically, up to corrections that are exponentially suppressed at large volume and large complex structure. By computing these corrections explicitly for carefully chosen fluxes, one can identify isolated vacua with exponentially small $W$, yielding Anti-de Sitter solutions with exponentially small cosmological constant \cite{Demirtas:2021nlu}, as well as fully specified candidate de Sitter vacua \cite{McAllister:2024lnt}.

    To better understand such solutions, and to investigate the structure of corrections beyond tree level, we continue the study of this class of flux vacua. Our main results are as follows.

    We argue that the \emph{quasi-moduli space} --- the asymptotic cone in field space where the superpotential is exponentially suppressed --- admits a smooth interpolation between weakly coupled O3/O7 orientifolds in type IIB, weakly coupled O6 orientifolds of $SU(3)$-structure manifolds in type IIA \cite{Grana:2004bg}, and weakly curved M-theory compactifications on $G_2$-manifolds.\footnote{See also \cite{VanHemelryck:2024bas} for related recent work in the case of weak $G_2$.} While the type IIB vacua involve both NSNS and RR fluxes, the type IIA models have only RR 2-form flux, and the M-theory duals are purely geometric. This implies that the PFV mechanism of \cite{Demirtas:2019sip} can be used to construct, at least implicitly, compact $G_2$-manifolds by solving algebraic (Diophantine) equations in the flux quanta. Though this procedure remains implicit, we hope it provides a new approach to studying and enriching the landscape of $G_2$-manifolds \cite{JoyceI,JoyceII,Kovavlev:2003,Corti:2012aa,Corti:2012kd,Halverson:2014tya,Braun:2017uku,Braun:2018fdp} and the resulting $4d$ $\mathcal{N}=1$ vacua in M-theory.

    To substantiate this proposal, in \S\ref{sec:toroidalPFVs} we analyze PFV-type flux vacua on a simple O3 orientifold of a six-torus. This model has the advantage of an explicit metric, allowing us to construct full ten-dimensional backgrounds from solutions to Poisson equations on the torus. Applying the Buscher rules along a suitable three-torus, following \cite{Kachru:2002sk,Grana:2006kf}, we obtain the mirror-dual type IIA backgrounds, including the warp factor and dilaton profiles --- again in terms of solutions to a Poisson equation, now on a three-torus. These backgrounds are twisted tori (nilmanifolds) with $SU(3)$-structure, previously studied in \cite{Kachru:2002sk,Grana:2006kf,Andriolo:2018yrz}, and carry only RR 2-form flux. They satisfy the criteria of \cite{Grana:2004bg} for lifting to fluxless $G_2$-compactifications in M-theory (or compactifications with further reduced holonomy). In a suitable approximation, we exhibit Ricci-flat $G_2$ metrics explicitly; these can be promoted to exact solutions by gluing in local copies of the Atiyah–Hitchin manifold \cite{Atiyah:1985fd}, in analogy with constructions of local K3 manifolds \cite{Sen:1997kz}. This yields explicit approximate metrics for K3, several Calabi–Yau threefolds, and a class of $G_2$-manifolds. While the toroidal flux backgrounds themselves are not new --- see \cite{Kachru:2002sk,Grana:2006kf,Andriolo:2018yrz,Cicoli:2022vny} --- our contribution is to reinterpret them as special cases of the PFVs in \cite{Demirtas:2019sip}, and to formulate the dualization steps in a way that facilitates generalization to PFVs of arbitrary Calabi–Yau orientifolds.

    This generalization is carried out in \S\ref{sec:GeneralCYcase}, where we map the PFV conditions in type IIB to corresponding constraints on fluxes and geometry in the dual type IIA models. By comparing to the classification of supersymmetric Minkowski vacua in \cite{Grana:2004bg}, we argue that the M-theory dual persists beyond the toroidal case. In \S\ref{sec:ParametricControl} we analyze the regimes of parametric control of the IIB, IIA, and M-theory descriptions and show that the vanishing superpotential limit of \cite{Demirtas:2019sip} corresponds to the universal decompactification limit in M-theory. This implies that the M-theory frame, being manifestly weakly coupled, is particularly well-suited for studying PFVs with fully stabilized moduli, such as the exponentially small cosmological constant vacua of \cite{Demirtas:2021nlu}.

    Finally, in \S\ref{sec:SingularBulk}, we revisit a potential control issue in warped type II flux compactifications \cite{Carta:2019rhx,Gao:2020xqh,Carta:2021lqg}, known as the singular bulk problem, which involves uncontrolled warping corrections to the Kähler potential and coordinates. We argue that this issue becomes parametrically worse at large complex structure, possibly leading to $\mathcal{O}(1)$ quantum corrections in vacua such as \cite{Demirtas:2021nlu,McAllister:2024lnt}. However, we point out that screening of D3-charge may mitigate this problem in models without large warped throats. Most importantly, we show that these corrections, while non-perturbative from the type II perspective, are fully classical in the M-theory frame --- and hence, in principle, computable if the dual $G_2$-manifold is known.\footnote{This geometrization of warping corrections has been emphasized in \cite{Andriolo:2018yrz} in the context of type IIA mirrors and M-theory uplifts of toroidal flux vacua.}

    Before turning to these results, we review GKP flux compactifications and the construction of PFVs via Diophantine equations in \S\ref{sec:PFVs}. We conclude with an outlook in \S\ref{sec:Conclusions}.

    \section{Perturbatively flat vacua in Type IIB string theory}\label{sec:PFVs}
    We will begin our discussion with a review of type IIB vacua with exponentially small superpotential, \emph{perturbatively flat vacua}, as introduced in \cite{Demirtas:2019sip}, and further discussed in \cite{Demirtas:2020ffz,Alvarez-Garcia:2020pxd,Demirtas:2021nlu,Honma:2021klo,Marchesano:2021gyv,Cicoli:2022vny,McAllister:2024lnt}. 
    
    \subsection{GKP flux vacua}\label{subsec:PFVs:GKP}
    We consider type IIB string theory on a Calabi-Yau orientifold $\tilde{X}/\mathbb{Z}_2$ of O3/O7 type. We will denote by $X$ the mirror-dual Calabi-Yau threefold. To avoid confusion, geometric quantities of $\tilde{X}$ will be marked correspondingly, e.g., $\tilde{h}^{p,q}$ denote the Hodge numbers of $\tilde{X}$, while $h^{p,q}$ denote those of $X$. We will also write the NSNS two form in type IIA(B) string theory as $B_2$ $(\widetilde{B}_2)$ respectively, with field strength $H_3$ $(\widetilde{H}_3)$.

	For simplicity, we will assume that the cohomology groups $H^{\text{even}}(\tilde{X},\mathbb{R})$ are all orientifold-even and that $H^3(\tilde{X},\mathbb{R})$ is orientifold-odd. 
	We will denote by $[\widetilde{\Sigma}_2]^i$, $i=1,\ldots,\tilde{h}^{1,1}$ a basis of $H^4(\widetilde{X},\mathbb{Z})$, and by $[\widetilde{\Sigma}_4]_i$ the dual basis of $H^2(\widetilde{X},\mathbb{Z})$. 
	
	The $4d$ $\mathcal{N}=1$ effective supergravity theory has $\tilde{h}^{1,1}+\tilde{h}^{2,1}+1$ chiral multiplets. Ignoring warping effects, the $\tilde{h}^{1,1}$ K\"ahler moduli can be defined from the RR fourform potential $C_4$ and the (string-frame) K\"ahler form $\tilde{J}$ on $\tilde{X}$ as
	\begin{equation}\label{eq:Kahler_moduli}
		T_i=\int_{\widetilde{X}}\left(\frac{1}{2}e^{-\phi_B}\tilde{J}\wedge \tilde{J}+iC_4\right) \wedge [\widetilde{\Sigma}_4]_i\, ,\quad i=1,\ldots,\tilde{h}^{1,1}\, ,
	\end{equation}
	where $e^{\phi_B}$ denotes the type IIB string coupling. The axio-dilaton is defined as $\tau:=C_0+ie^{-\phi_B}$ with $C_0$ the RR zero form.
	
	Complex structure moduli space has complex dimension $\tilde{h}^{2,1}$, and natural ``flat'' coordinates are given by a subset of the periods of the holomorphic three-form $\tilde{\Omega}$. Letting $\{[\widetilde{\Sigma}_3]_A,[\widetilde{\Sigma}_3]^A\}$ be a symplectic basis of $H^3(\widetilde{X},\mathbb{Z})$, i.e., 
	\begin{equation}\label{eq:symplectic_basis_IIB}
		\int_{\tilde{X}} [\widetilde{\Sigma}_3]_A\wedge [\widetilde{\Sigma}_3]^B=\delta_A^B\, ,\quad \int_{\widetilde{X}} [\widetilde{\Sigma}_3]_A\wedge [\widetilde{\Sigma}_3]_B=\int_{\tilde{X}} [\widetilde{\Sigma}_3]^A\wedge [\widetilde{\Sigma}_3]^B=0\, ,\quad A=0,\ldots,\tilde{h}^{2,1}\, ,
	\end{equation} 
	one can define periods $z^A:=\int_X \widetilde{\Omega}\wedge [\widetilde{\Sigma}_3]^A$ and  $\mathcal{F}_A:=\int_X \widetilde{\Omega}\wedge [\widetilde{\Sigma}_3]_A$, and use the $z^A$ as local projective coordinates on complex structure moduli space. The $\mathcal{F}_A$ are then determined by the $z^A$ through a prepotential $\mathcal{F}(z^A)$:
	\begin{equation}
		\mathcal{F}_A=\del_{z^A}\mathcal{F}\, ,
	\end{equation}
	with $\mathcal{F}$ homogeneous of degree two. 
	
	Away from $z^0=0$ we may set $z^0=1$ and use the remaining $z^a$, $a=1,\ldots,\tilde{h}^{2,1}$ as local coordinates on moduli space. Assuming an appropriately chosen basis of three-cycles, the coordinates $z^a$ are equivalent to the complexified K\"ahler moduli of type IIA on the mirror threefold $X$.
	
	The prepotential $\mathcal{F}$, in the gauge $z^0=1$, can be decomposed into a polynomial part $\mathcal{F}_{\text{pol.}}$ and a series of exponential corrections $\mathcal{F}_{\text{inst.}}$ according to \cite{Candelas:1990rm,Candelas:1993dm,Hosono:1993qy,Candelas:1994hw,Hosono:1994ax}
	\begin{align}\label{eq:prepotential}
		\mathcal{F}(z)&=\mathcal{F}_{\text{pol.}}(z)+\mathcal{F}_{\text{inst.}}(z)\, ,\\
		\mathcal{F}_{\text{pol.}}(z)&=-\frac{1}{3!}\kappa_{abc}z^az^bz^c+\frac{1}{4}\mathbb{A}_{ab}z^az^b+\frac{1}{24}c_a z^a+\frac{\chi(X) \zeta(3)}{2(2\pi i)^3}\, ,\\
		\mathcal{F}_{\text{inst.}}(z)&=-\frac{1}{(2\pi i)^3}\sum_{\beta}n^0_\beta \text{Li}_3(e^{2\pi i \langle \beta, z \rangle})\, .
	\end{align}
	The limit in complex structure moduli space in which all exponential corrections go to zero is called Large Complex Structure (LCS), and is the mirror dual of the large volume limit of $X$. Accordingly, the integral tensors $\kappa,\mathbb{A},c$ and $\chi(X)$ are given by classical geometric data of the mirror threefold $X$ and will be defined precisely in \S\ref{subsec:GeneralCYcase:IIA_flux_quantization}. The $n_\beta^0$ are the genus zero Gopakumar-Vafa invariants of $X$ \cite{Gopakumar:1998ii,Gopakumar:1998jq}, the sum runs over effective curve classes $\beta\in H_2(X,\mathbb{Z})$, and Li$_n(q):=\sum_{k=1}^\infty q^k/k^n$.
	
	Turning on three-form fluxes $F_3$ and $H_3$ generates a superpotential in the four-dimensional effective theory, which is given by \cite{Gukov:1999ya}
	\begin{equation}
		W=\int_X (F_3-\tau \widetilde{H}_3)\wedge \widetilde{\Omega}=W^{RR}(z)+W^{NSNS}(\tau,z)\, ,
	\end{equation}
	where $\widetilde{\Omega}$ denotes the holomorphic three-form on $\widetilde{X}$.
	
	Using the LCS expansion \eqref{eq:prepotential} of the prepotential, and defining quantized fluxes
	\begin{align}\label{eq:threeform_fluxes}
		f^A:&=\int_{\widetilde{X}}F_3\wedge [\widetilde{\Sigma}_3]^A\, ,\quad f_A:=\int_{\widetilde{X}}F_3\wedge [\widetilde{\Sigma}_3]_A\, ,\\ h^A:&=\int_{\widetilde{X}}\widetilde{H}_3\wedge [\widetilde{\Sigma}_3]^A\, ,\quad h_A:=\int_{\widetilde{X}}\widetilde{H}_3\wedge [\widetilde{\Sigma}_3]_A\, ,
	\end{align}
	we may expand the RR superpotential $W^{RR}(z)$ around LCS. 
	
	Setting $f^A=(-m^0,m^a)$ and $f_A=(\tilde{e}_0,\tilde{e}_a)$, one finds
	\begin{align}\label{eq:RR_superpotential}
		&W^{RR}(z)=W^{RR}_0(z)+\delta W^{RR}(z)\, ,\\
		&W^{RR}_0=m^0\frac{1}{3!}\kappa_{abc}z^az^bz^c+m^a\frac{1}{2}\kappa_{abc}z^b z^c+e_a z^a+e_0\, ,\\
		&\delta W^{RR}=-\frac{\zeta(3)\chi}{(2\pi i)^3}m^0+\sum_{\beta} n_\beta^0\left( (m^a+m^0 z^a)\beta_a \frac{\text{Li}_2(e^{2\pi i \langle \beta,z \rangle})}{(2\pi i)^2}-2m^0 \frac{\text{Li}_3(e^{2\pi i \langle \beta,z \rangle})}{(2\pi i)^3}\right)\, .
	\end{align}
	The NSNS superpotential is of course similarly expanded in terms of the $h^A$ and $h_A$.
	
	For later convenience, we have written the polynomial part of the superpotential in terms of shifted, and in general non-integer, flux numbers
	\begin{equation}\label{eq:shifted_fluxes}
		e_a:=\tilde{e}_a+\frac{1}{2}\mathbb{A}_{ab}m^b-\frac{c_a}{24}m^0\, ,\quad e_0:=\tilde{e}_0-\frac{c_a}{24}m^a\, .
	\end{equation}
	The integer quantized three form fluxes $(F_3,H_3)$ induce D3-brane charge in $\tilde{X}$ due to the Bianchi identity of the self-dual five-form field strength,
	\begin{equation}\label{eq:RR5_Bianchi}
		dF_5=\widetilde{H}_3\wedge F_3+\text{localized sources}\, ,
	\end{equation}
	where the localized sources that contribute to the r.h. side are D3 branes and O3 planes, as well as D7 branes, and O7 planes via their induced D3-brane charge. For a configuration of D7 branes that cancels the D7-tadpole locally --- i.e., for 4 D7-branes on top of each O7 plane --- the overall charge cancellation condition reads
	\begin{equation}\label{eq:D3_tadpole}
		Q+N=\mathcal{Q}\, ,\quad 
		Q:=\frac{1}{2}\int_{\tilde{X}}\widetilde{H}_3\wedge F_3\, ,\quad \mathcal{Q}:=\frac{\chi(O)}{4}\, ,
	\end{equation}
	where $O$ denotes the orientifold fixed point locus, and $\chi$ is the topological Euler characteristic. $N\geq 0$ denotes the number of mobile D3 branes.
	
	Any solution to the F-term equations $D_\tau W=D_{z^a}W=0$ corresponds to a warped ten dimensional supergravity solution of GKP type \cite{Giddings:2001yu}, with constant dilaton $e^{-\phi_B}$ and ten dimensional (string frame) metric
	\begin{equation}
		ds^2=e^{2A}dx^2+e^{-2A}ds^2_{\text{CY}}\, ,
	\end{equation}
	where the internal metric $ds^2_{\text{CY}}$ remains Ricci flat, up to the conformal factor $e^{-2A}$, and the complexified three-form $G_3=F_3-\tau \widetilde{H}_3$ is imaginary self-dual $*_6 G_3=i G_3$.
	
	Due to four dimensional Poincar\'e invariance, the five form field strength takes the form $F_5=(1+*)d\alpha \wedge d^4x$, for $\alpha$ some function on $\widetilde{X}$. The Bianchi identity \eqref{eq:RR5_Bianchi} and Einstein's equations are solved if $\alpha=e^{4A}$, and if the warp factor satisfies \cite{Giddings:2001yu,Giddings:2005ff}
	\begin{equation}\label{eq:6delectrostatic}
		e^{-\phi_B}\nabla^2_{\widetilde{X}}e^{-4A} =\rho_{\text{D3}}^{\text{flux}}+\rho^{\text{loc}}_{\text{D3}}\, ,\quad \text{with}\quad  \rho_{\text{D3}}^{\text{flux}}=\frac{e^{\phi_B}}{12}(G_3)_{i jk}(\overline{G_3})^{ijk}\, ,
	\end{equation}
	and with $\rho^{\text{loc}}_{\text{D3}}$ the D3-charge density of localized sources. In toroidal compactifications, considered in \S\ref{sec:toroidalPFVs},  one simply has $\rho_{\text{D3}}^{\text{flux}}=2Q/\widetilde{\mathcal{V}}$.\footnote{An earlier version erroneously contained this simplified expression for $\rho_{\text{D3}}^{\text{flux}}$ already in \eqref{eq:6delectrostatic}. I thank Severin L\"ust for pointing this out to me.} In \eqref{eq:6delectrostatic} indices are raised with the metric $ds^2_{\text{CY}}$, and we denote by $\tilde{\mathcal{V}}$ the volume of $\widetilde{X}$, as measured with the metric $ds^2_{CY}$. Explicitly, one has
	\begin{equation}\label{eq:RR5_GKP}
		F_5=d e^{4A} \wedge d^4x+*_{\text{CY}} d e^{-4A}=d e^{4A} \wedge d^4x-4*_{\text{int}} d A\, ,
	\end{equation}
	where $*_{\text{CY}}$ denotes the Hodge star w.r.t. the metric $ds^2_{\text{CY}}$, and $*_{\text{int}}$ is the Hodge star with respect to the physical internal metric $e^{-2A}ds^2_{CY}$.

	\subsection{Perturbatively flat vacua}\label{subsec:PFVs:PFVs}
	We are now ready to define a \emph{perturbatively flat vacuum} (PFV) following \cite{Demirtas:2019sip}. First, we set $m^0=0$, and choose $m^a$ subject to the integrality conditions
	\begin{equation}\label{eq:integrality_conditions}
		 \mathbb{A}_{ab}m^b\equiv 0\mod 2\, ,\quad 
		 c_a m^a=0\mod 24.
	\end{equation}
	In this case the shifted flux numbers $(e_a,e_0)$ satisfy integer quantization conditions. This allows us to set $e_a=e_0=0$. We further set 
	\begin{equation}\label{eq:PFV_Hflux}
		h^A=0\, ,\quad  \text{and}\quad  h_A=(0,k_a)\, ,
	\end{equation}
	for some integral vector $\vec{k}$ subject to the following condition
	\begin{equation}\label{eq:PFV_Diophantine_eq}
		\exists\, \vec{p}\in \mathcal{K}_X\,:\quad \langle \vec{k},\vec{p}\rangle = 0\, ,\quad \text{and} \quad N_{ab}p^b=k_a\, ,
	\end{equation}
	with $N_{ab}:=\kappa_{abc}m^c$. Here, $\mathcal{K}_X$ denotes the K\"ahler cone of the mirror dual threefold $X$.
	
	If these conditions are satisfied the superpotential simplifies dramatically,
	\begin{align}\label{eq:PFV_LCS_corrections}
		W(z,\tau)\quad &\longrightarrow \quad W_{\text{pert.}}(z,\tau) +\delta W_{\text{inst.}}(z)\, ,\\
		W_{\text{pert.}}(z,\tau)&:=\frac{1}{2}N_{ab}z^a z^b-\tau k_a z^a\, ,\\
		\delta W_{\text{inst.}}(z)&:=-\frac{1}{(2\pi )^2}\sum_{\beta} n_\beta^0 m^a\beta_a \text{Li}_2(e^{2\pi i \langle \beta,z \rangle})\, .
	\end{align}
	Crucially, upon neglecting the exponential terms in $\delta W_{\text{inst,}}$ all F-term conditions are solved with vanishing superpotential along the sub-locus of moduli space where 
	\begin{equation}\label{eq:PFV_locus}
		z^a=p^a\tau\, ,
	\end{equation}
    and the corrections $\delta W_{\text{inst.}}(z)$ decay exponentially at large $\text{Im}(\tau)$. In the generic case where $\det N\neq 0$ the condition \eqref{eq:PFV_Diophantine_eq} is equivalent to imposing a Diophantine equation in flux quanta
    \begin{equation}
        N^{ab}k_a k_b\overset{!}{=} 0\, ,\quad N^{ab}:=(N^{-1})^{ab}\, ,
    \end{equation}
    as well as the K\"ahler cone condition $N^{ab}k_b\in \mathcal{K}_X$.

	Integrating out the complex structure deformations that don't satisfy \eqref{eq:PFV_locus} one ends up with a low energy effective supergravity theory with $\tilde{h}^{1,1}+1$ chiral multiplets, namely the K\"ahler moduli $T_i$ introduced in \eqref{eq:Kahler_moduli} and one further multiplet whose complex scalar component parameterizes the joint perturbations of the $z^a$ and $\tau$ that maintain the relation \eqref{eq:PFV_locus}.\footnote{Further moduli can exist if the choice of $\vec{p}$ in \eqref{eq:PFV_locus} is not unique, i.e., if $\det N=0$.} Without loss of generality we may parameterize this locus by the axio dilaton $\tau$.

    The effective superpotential will in general receive corrections from euclidean D3-instantons as well as strong IR gauge dynamics on seven-brane gauge theories, which are non-perturbative in the $T_i$ \cite{Witten:1996bn}. Together with the exponential corrections $\delta W_{\text{inst.}}(z)$ these have been used in \cite{Demirtas:2021nlu} to stabilize all moduli in Anti-de Sitter vacua with extremely small vacuum energy. In these vacua, the vevs of the $\tilde{h}^{1,1}+1$ moduli $(\tau,T_i)$ scale as
    \begin{equation}
        \langle \text{Re}(T_i)\rangle \sim \langle \text{Im}(\tau)\rangle\sim \lambda:=\frac{\log(1/|W_0|)}{2\pi}\gg 1\, ,
    \end{equation}
    where $W_0$ is the vev of the superpotential.

\section{Toroidal Models}\label{sec:toroidalPFVs}

    \subsection{An O3 orientifold of $T^6$}\label{subsec:toroidalPFVs:O3orientifold}
    As a concrete example, we consider a classic type IIB O3 orientifold of $T^6$. We denote by $\vec{X}=(\tilde{x}^1,y_1,\tilde{x}^2,y_2,\tilde{x}^3,y_3)$ six torus coordinates, with unit periodicities, and define an O3 orientifold via the involution $\vec{X}\mapsto -\vec{X}$, leading to $2^6=64$ O3-planes as, e.g., in \cite{Frey:2002hf,Kachru:2002sk,Cicoli:2022vny}. We take all $O3$-planes to be the standard $O3^-$-planes, and thus the overall D3-charge of the O-planes is $-64/4=-16$ and needs to be canceled via a combination of background fluxes and mobile D3-branes.

    With $16$ mobile D3-branes and no fluxes, the four-dimensional effective theory is $\mathcal{N}=4$ supergravity, with $22$ vector multiplets, and is T-dual to the type I theory on $T^6$. The massless bosonic degrees of freedom in the closed string sector arise as follows:
    \begin{itemize}
        \item From the 10d metric $g_{MN}$ we get the $4d$ metric and 21 real scalars.
        \item The 10d dilaton gives one real scalar in $4d$.
        \item $B_2$ yields six abelian vector fields: $B_2=\mathcal{B}_{I\mu} dx^\mu\wedge dX^I$.
        \item $C_0$ is one real scalar.
        \item $C_2$ yields six abelian vector fields: $C_2=\mathcal{C}_{I\mu} dx^\mu\wedge dX^I$.
        \item $C_4$ yields 15 real scalars: $C_4=c_\alpha(x^\mu) \omega^\alpha$ with $\omega^\alpha$ a basis of harmonic fourforms. 
    \end{itemize}
    In addition there are $16$ vector multiplets in the open string sector.
    
    Due to the orientifold involution, the flat internal part of the type IIB $B$-field is frozen to take values in $H^2(T^6,\mathbb{Z}_2)$ and we will set it to zero.
    
    In total the closed string moduli parameterize a 38 dimensional moduli space. It will be convenient to view the six torus as a K\"ahler threefold parameterized by three complex coordinates $\tilde{\phi}^a:=\tilde{x}^a+z^{ab} y_b$, written in terms of nine complex structure parameters $z^{ab}$. The Riemannian metric on $T^6$ is then specified by a hermitian metric $g_{a\bar{b}}$, or equivalently a choice of K\"ahler form
    \begin{equation}
        \widetilde{J}=\frac{i}{2} g_{a\bar{b}}d\tilde{\phi}^a\wedge d\overline{\tilde{\phi}}^{\bar{b}}\, ,
    \end{equation}
    and the holomorphic three-form is 
    \begin{equation}
    \widetilde{\Omega}=d\tilde{\phi}^1\wedge d\tilde{\phi}^2 \wedge d\tilde{\phi}^3\, .
    \end{equation}
    The 21 metric moduli can be thought of as the 9 real degrees of freedom of the hermitian metric $g_{a\bar{b}}$ and the 9 complex parameters $z^{ab}$ modulo the complex structure deformations that change the holomorphic three-form by $J\wedge \delta$ with $\delta$ a closed (1,0)-form. Explicitly, for fixed $g_{a\bar{b}}$, we have an equivalence relation on small complex structure deformations $z\rightarrow z+\delta z$
    \begin{equation}
        (\delta z)^{ab} \sim (\delta z)^{ab}+\epsilon^{acd}g_{d\bar{e}}{\text{Im}(z)^{\bar{e}b}} \delta_c\, ,\quad \text{with} \quad \vec{\delta}\in \mathbb{C}^3\, .
    \end{equation}
    
    Thus, only six combinations of the nine complex structure deformations are physical, and we indeed end up with $9+2\cdot 6=21$ real metric deformations.

    \subsection{Three-form fluxes}\label{subsec:toroidalPFVs:fluxes}
    For generic internal three-form fluxes $F_3$ and $H_3$ supersymmetry gets broken to $4d$ $\mathcal{N}=1$, which generically is further broken to nothing by the F-terms of the remaining chiral multiplets. 
    
    The chiral multiplets are constructed as follows \cite{Kachru:2002sk}: the 15 $C_4$ axions receive St\"uckelberg kinetic terms
    \begin{align}
        F_5\supset &\del_\mu c_{\alpha} dx^\mu \wedge \omega^\alpha+\frac{1}{2}\mathcal{B}_I^\mu dx^\mu \wedge dX^I\wedge F_3-\frac{1}{2}\mathcal{C}_I^\mu dx^\mu \wedge dX^I\wedge H_3\\
        &= \left(\del_{\mu}c_\alpha -\mathcal{B}_{\mu I} {P^I}_\alpha-\mathcal{C}_{\mu I} {Q^I}_\alpha\right)\wedge \omega^\alpha\, ,
    \end{align}
    with St\"uckelberg charge matrices
    \begin{equation}
        -\frac{1}{2}dX^I\wedge F_3\equiv {P^I}_\alpha \omega^\alpha\, ,\quad \frac{1}{2}dX^I\wedge H_3\equiv {Q^I}_\alpha \omega^\alpha \, .
    \end{equation}
    Thus, generically, twelve combinations of the $c_\alpha$ get eaten by the 12 abelian vectorfields such that only three gauge-invariant $C_4$ axions and no vectorfields remain in the massless spectrum (except the open string $U(1)$s). Furthermore, the massive $U(1)$ gauge fields lead to D-term constraints on the K\"ahler class
    \begin{equation}
        \widetilde{J}\wedge F_3=\widetilde{J}\wedge H_3=0\, ,
    \end{equation}
    which generically make six K\"ahler parameters massive. The remaining three pair up with the three massless $C_4$ axions to form the bosonic components of three chiral multiplets.

    But, for non-generic fluxes, some of the abelian vector fields can remain massless, leading to additional massless $C_4$-axions, and extra D-flat K\"ahler moduli may arise as well. Such configurations may preserve $\mathcal{N}=2$ or more supersymmetry in four dimensions, as in \cite{Frey:2002hf,Kachru:2002sk}. All explicit solutions we will construct in this paper will feature such additional massless degrees of freedom.

	We will restrict the discussion to a family of six tori, which we take to be the product of three copies of $T^2$. Thus, the (string-frame) metric is
    \begin{equation}
        ds^2=\sum_{a=1}^3 \tilde{t}^a\frac{d\tilde{\phi}^a d\bar{\tilde{\phi}}^{\bar{a}}}{t^a}=\sum_a \tilde{t}^a\left(\frac{1}{t^a}(\tilde{w}^a)^2+t^a(dy^a)^2\right)
    \end{equation}
    corresponding to diagonal complex torus coordinates $\tilde{\phi}^a:=\tilde{x}_a+z^a y_a$, and we define one-forms 
    \begin{equation}
        \tilde{w}^a:=d\tilde{x}_a+b^a dy_a\, .
    \end{equation}
    The diagonal metric ansatz is parameterized in terms of three complex structure moduli $z^a\equiv b^a+i t^a$ and three K\"ahler moduli $\tilde{t}^a$, 
    \begin{equation}\label{eq:IIB_kahler_class}
        \widetilde{J}=\sum_a \tilde{t}^a \frac{i}{2t^a}d\tilde{\phi}^a d\bar{\tilde{\phi}}^a=\sum_a \tilde{t}^a d\tilde{x}^a \wedge dy^a\, .
    \end{equation}
    These moduli are all physical and parameterize a real nine-dimensional sub-locus of metric moduli space.
    The (string-frame) six-torus volume is equal to
    \begin{equation}
        \widetilde{\mathcal{V}}:=\tilde{t}^1\tilde{t}^2 \tilde{t}^3\, .
    \end{equation}
    The relevant periods of the holomorphic three-form are
    \begin{equation}
        \Pi_A= \int \widetilde{\Omega}\wedge \vec{\gamma} = \begin{pmatrix}
            \mathcal{F}_0\\
            \mathcal{F}_a\\
            1\\
            z^a
        \end{pmatrix} \, ,\quad \mathcal{F}_a:=\del_{z^a}\mathcal{F}\, ,\quad \mathcal{F}_0:=2\mathcal{F}-z^a\mathcal{F}_a
    \end{equation}
    in terms of a prepotential 
    \begin{equation}
        \mathcal{F}(z):=-\frac{1}{3!}\kappa_{abc}z^a z^b z^c:= -z^1 z^2 z^3\, ,
    \end{equation}
    when expanded in the following set of three-forms
    \begin{align}
       \vec{\gamma} =\begin{pmatrix}
           \beta_0\\
           \beta_a\\
           \alpha^0\\
           \alpha^a
       \end{pmatrix}\, ,\quad \alpha^a=&  \begin{pmatrix}
            -d\tilde{x}_1 \wedge dy_2 \wedge dy_3 \\
            -dy_1 \wedge d\tilde{x}_2 \wedge dy_3 \\
            -dy_1 \wedge dy_2 \wedge d\tilde{x}_3
        \end{pmatrix}\, ,\quad \beta_a=\begin{pmatrix}
            dy_1\wedge d\tilde{x}_2 \wedge d\tilde{x}_3 \\
            d\tilde{x}_1\wedge dy_2 \wedge d\tilde{x}_3 \\ 
            d\tilde{x}_1\wedge dx_2 \wedge dy_3
        \end{pmatrix}\, ,\\
        \alpha_0=& -dy_1 \wedge dy_2 \wedge dy_3\, ,\quad \beta_0=-d\tilde{x}_1\wedge d\tilde{x}_2 \wedge d\tilde{x}_3\, .
    \end{align}
    These forms satisfy the standard symplectic pairing relations
    \begin{equation}
        \int_{\tilde{X}}\alpha^A\wedge \beta_B={\delta^A}_B\, ,\quad  \alpha^A\wedge \alpha^B=\beta_A\wedge \beta_B=0\, ,
    \end{equation}
    and their periods suffice to parameterize our diagonal torus $T^6=T^2\times T^2\times T^2$.

    We now make a restricted ansatz for fluxes as in \cite{Demirtas:2019sip},
    \begin{equation}\label{eq:three_form_torusPFV}
        F_3=-m^a\beta_a\, ,\quad H_3=k_a\alpha^a\, ,
    \end{equation}
    in terms of flux quanta $\vec{m},\vec{k}\in (2\mathbb{Z})^3$.\footnote{We take fluxes to be even in the covering space $T^6$ such that all O3 planes can consistently be taken to be of $O3^-$ type, \emph{cf.} the discussion of \cite{Frey:2002hf}.}

    These fluxes induce 
    \begin{equation}
        Q=-\frac{1}{2}\langle \vec{m},\vec{k}\rangle
    \end{equation}
    units of D3-brane charge. 
    
    We will further assume that the three-form flux quanta satisfy the PFV conditions \S\ref{subsec:PFVs:PFVs} (and we will construct explicit such solutions in \S\ref{subsec:toroidalPFVs:concrete_solutions}). I.e., we demand that there exists a $\vec{p}\in \mathbb{R}_+^3$ such that
    \begin{equation}
        \langle \vec{k},\vec{p}\rangle=0\quad \text{and}\quad  N\cdot \vec{p}=\vec{k}\, ,\quad N_{ab}:=\kappa_{abc}m^c=\begin{pmatrix}
            0 & m^3 & m^2\\
            m^3 & 0 & m^1\\
            m^2 & m^1 & 0
        \end{pmatrix}\, .
    \end{equation}
    In this case, the superpotential and its F-terms vanish identically for
    \begin{equation}
        \vec{z}=\vec{p}\tau\, ,
    \end{equation}
    and we will henceforth parameterize this locus by the axio-dilaton $\tau \equiv b+i t$.
    
    Our flux choice is automatically primitive for diagonal K\"ahler class \eqref{eq:IIB_kahler_class}, and the total D3-charge vanishes in the presence of 
    \begin{equation}
        N:=16-Q
    \end{equation}
    mobile D3-branes (and their orientifold images).

    \subsection{10d type IIB solution}\label{subsec:toroidalPFVs:IIBsolution}
    We will begin by studying the ten-dimensional type IIB supergravity solution, a particular GKP solution \cite{Giddings:2001yu}.
    
    First, the RR five-form is given implicitly via \eqref{eq:RR5_GKP} in terms of the warp factor $e^{2A}$, which itself is a solution to the electro-static problem on $T^6$,
    \begin{equation}
        \tilde{\mathcal{V}}e^{-\phi_B}\nabla^2_{T^6}e^{-4A}=2Q+\sum_{\alpha=1}^{2N}\delta^{(6)}(\tilde{\phi}-\tilde{\phi}_\alpha^{D3})-\frac{1}{2}\sum_{r=1}^{64}\delta^{(6)}(\tilde{\phi}-\tilde{\phi}_r^{O3})\, .
    \end{equation}
    
    Correspondingly \cite{Giddings:2001yu}, the string-frame metric of the six-torus gets conformally rescaled and the ten-dimensional metric becomes 
    \begin{equation}
        ds^2=e^{2A}dx^\mu dx_\mu +e^{-2A}\sum_a \tilde{t}^a\left(\frac{1}{p^a t}(\tilde{w}^a)^2+p^at (dy^a)^2\right)
    \end{equation}
    The three form field strengths \eqref{eq:three_form_torusPFV} solve the equations of motion and Bianchi identity, and the type IIB $B$-field can locally be written as
    \begin{equation}
        \tilde{B}=\sum_a A^a\wedge d\tilde{x}^a\, ,\quad A^a=-\begin{pmatrix}
            k_1y^2 dy^3\\
            k_2y^3 dy^1\\
            k_3y^1 dy^2
        \end{pmatrix}\, .
    \end{equation}
    If $t\gg 1$ the string coupling is small, and if $\tilde{t}^a\gg t$ all string-frame cycle volumes are large. In this regime the type IIB description is parametrically controlled.
    
    It will be useful to view our six-torus as a (trivial) $T^3$-fibration over a base $T^3$, where the base is parameterized by the $\vec{y}$-coordinates and the fiber is parameterized by the $\vec{\tilde{x}}$-coordinates. We will refer to the fiber torus as the SYZ-fiber, by analogy to the general Calabi-Yau case \cite{Strominger:1996it}. The requirement $\tilde{t}^a\gg t$ then ensures that the SYZ-fiber has large string-frame volume.

    \subsection{Mirror-dual type IIA solution}\label{subsec:toroidalPFVs:IIAsolution}
    The type IIB solution admits various T-dual descriptions, but we will be particularly interested in the ``mirror-dual'' type IIA description, which we obtain by performing three T-dualities along the SYZ-fiber directions $(\tilde{x}^1,\tilde{x}^2,\tilde{x}^3)$. Various aspects of the resulting class of solutions have previously been studied e.g. in \cite{Kachru:2002sk,Gurrieri:2002wz,Grana:2006kf,Caviezel:2008ik,Andriolo:2018yrz}, and they turn out to fall into a category of SUSY-preserving $SU(3)$-structure backgrounds classified in \cite{Kaste:2003dh,Grana:2004bg,Behrndt:2004mj}.\footnote{See \cite{Grana:2005jc} for a review of this subject.}
    
    The O3-planes and D3-branes in the type IIB description get mapped to O6-planes and D6-branes in the type IIA picture. These  sit at points in the base torus, while wrapping the entire SYZ-fiber. Thus, fiber-wise, we expect to restore a perfect geometric shift-symmetry along the fiber directions, and so we expect to be able to obtain this dual type IIA background in the following three steps:
    \begin{enumerate}
        \item As in \cite{Grana:2004bg}, smear D-brane and O-plane sources along the fiber-direction, thus restoring three  $U(1)$-isometries in the type IIB background.
        \item Apply the Buscher rules to the NSNS background.
        \item T-dualize RR field strengths as usual.
    \end{enumerate}
    We emphasize that the smearing procedure (a) is just a technical trick to construct in simple terms the dual type IIA description of these vacua. The resulting backgrounds will solve the equations of motion and Bianchi identities with fully localized sources.
    
    After (a) the warp factor is independent of the $\tilde{x}$-coordinates and it will be convenient to write it as
    \begin{equation}
        e^{-4A}
        =\frac{\mathcal{V}_p^{\frac{1}{3}}}{\tilde{\mathcal{V}}^{\frac{2}{3}}}\chi(y;y^{D6})\, ,\quad \mathcal{V}_p:=\frac{1}{3!}\kappa_{abc}p^a p^b p^c=p^1 p^2 p^3\, .
    \end{equation}
    with $\chi(y;y^{D6})$ a solution to the 3d electro-static problem
     \begin{equation}\label{eq:3d_electrostatic}
        \hat{\nabla}^2 \chi(y;y^{D3})=2Q+\sum_{\alpha=1}^{2N}\delta^{(3)}(y-y_\alpha^{D3})-4\sum_{p=1}^{8}\delta^{(3)}(y-y_p^{O3})\, .
    \end{equation}
    Here, the last sum runs over the eight inequivalent O3-plane locations in the base $T^3_{y}$ (which will become the O6-plane positions), and $\hat{\nabla}^2$
    is the Laplacian on a \emph{unit-volume} base torus with metric 
    \begin{equation}\label{eq:3d_unit_volume_metric}
        d\hat{s}^2=\sum_a \hat{p}^a\xi^a dy_a^2\, ,\quad \xi^a:=\tilde{t}^a/\tilde{\mathcal{V}}^{\frac{1}{3}}\, ,\quad \hat{p}^a:=p^a/\mathcal{V}_p^{\frac{1}{3}}\, .
    \end{equation}
    The general solution $\chi(y;y^{D3})$ can be written as
    \begin{equation}\label{eq:chi_solution}
        \chi=\chi_0+c\, ,\quad \int d^3y \,\chi_0(y;y^{D3})=0\, ,
    \end{equation}
    in terms of an unconstrained modulus $c\in \mathbb{R}$.
    
    Using this solution we can write the type IIB RR five-form as
    \begin{equation}
        F_5|_{\text{internal}}= \tilde{w}^1\wedge \tilde{w}^2\wedge \tilde{w}^3 \wedge \hat{*} d\chi \, ,
    \end{equation}
    with $\hat{*}$ the Hodge star operator with respect to $d\hat{s}^2$.

    \subsubsection{NSNS sector}\label{subsubsec:toroidalPFVs:IIAsolution:NSNSsector}
    Applying the Buscher rules yields a type IIA O6 orientifold of a (twisted) six-torus that we will denote by $X_k$, with metric
    \begin{equation}\label{eq:twisted_torus}
        ds^2=t\sum_a p^a \left(r_{ab}w^aw^b+r^{ab} dy_a dy_b\right)\, ,\quad w^a:=dx^a+A^a\, ,
    \end{equation}
     where $r^{ab}:=e^{-2A} \tilde{t}^a \delta^{ab}=\mathcal{V}_p^{\frac{1}{6}}\sqrt{\chi}\xi^a \delta^{ab}$ and $r_{ab}:=(r^{-1})_{ab}$, and with orientifold involution defined by
    \begin{equation}
        \vec{y}\mapsto -\vec{y}\, .
    \end{equation}
    This manifold is a twisted $T_{\vec{x}}^3$-fibration over the base $T^3_{\vec{y}}$, with twist defined by the monodromies
    \begin{align}\label{eq:monodromies}
        I_1:&\, (y_2,x^1)\simeq (y_2+1,x^1+k_1y_3)\, ,\\
        I_2:&\, (y_3,x^2)\simeq (y_3+1,x^2+k_2y_1)\, ,\\
        I_3:&\, (y_1,x^3)\simeq (y_1+1,x^3+k_3y_2)\, .
    \end{align}
    As a consequence, while there are still two sets of globally defined one-forms, $dy^a$ and $w^a$, the twist implies that the one-forms $w^a$ are not closed:
    \begin{equation}
        d\vec{w}=-\begin{pmatrix}
            k_1 dy^2\wedge dy^3\\
            k_2 dy^3\wedge dy^1\\
            k_3 dy^1\wedge dy^2
        \end{pmatrix}\, .
    \end{equation}
    Furthermore, the type IIA dilaton $\phi_A$ acquires a non-trivial profile but this profile is completely determined by the warp factor:
    \begin{equation}
        e^{\phi_A}=e^{\phi_B}\text{Vol}(T^3_{\vec{x}})\equiv \sqrt{\frac{t\mathcal{V}_p}{\tilde{\mathcal{V}}}}e^{3A}=:\alpha \,e^{3A}\, .
    \end{equation}
    The manifold \eqref{eq:twisted_torus} has $SU(3)$ structure. To see this, we first define holomorphic and anti-holomorphic one forms
    \begin{equation}\label{eq:hol_oneforms}
        \eta^a:=w^a+i r^{ab} dy_b\, ,\quad \overline{\eta}^{\bar{a}}:=w^a-ir^{ab} dy_b\, .
    \end{equation}
    The dual basis of vector fields is then given by
    \begin{equation}\label{eq:hol_vectorfields}
    \del_a:=\frac{1}{2}\frac{\del}{\del x^a}+\frac{i}{2} r_{ab}\left(A^{bc}\frac{\del}{\del x^c}+\frac{\del}{\del y_b}\right)\, ,\quad \bar{\del}_{\bar{a}}:=\frac{1}{2}\frac{\del}{\del x^a}-\frac{i}{2} r_{ab}\left(A^{bc}\frac{\del}{\del x^c}+\frac{\del}{\del y_b}\right) \, ,
    \end{equation}
    where $A^a\equiv A^{ab}dy_b$. These vector fields define an \emph{almost complex structure} on $X_k$, and as usual, the exterior derivative
    decomposes as $d=\del +\bar{\del}$ with
    \begin{equation}
        \del:=\sum_a \eta^a  \del_a \, ,\quad \bar{\del}:=\sum_a \bar{\eta}^{\bar{a}} \bar{\del}_{\bar{a}}\, .
    \end{equation}
    We note that for a function of the base coordinates $y_a$, such as the warp factor $A=A(y)$ we get
    \begin{equation}
        dA=r_{ab}\del_{y_b}A \times \frac{i}{2}(\eta^a-\bar{\eta}^{\bar{a}})\, ,\quad \del_a A=-\bar{\del}_{\bar{a}}A\, .
    \end{equation}
    The O6 orientifold is anti-holomorphic with respect to the almost complex structure, in the sense that it maps
    \begin{equation}
        \eta^a\mapsto \overline{\eta^a}\, .
    \end{equation}
    
    With respect to our basis of one-form fields the metric \eqref{eq:twisted_torus} is hermitian, 
    \begin{equation}
        ds^2=g_{a\bar{b}}\eta^a\otimes \overline{\eta^b}\, ,\quad g_{a\bar{b}}:=t \,p^a r_{ab}\, ,
    \end{equation}
    and we can construct a globally defined real $(1,1)$-form $J$ and a complex $(3,0)$ form $\Omega$,
    \begin{align}
        J:=&\frac{i}{2}\sum_a g_{a\bar{b}}\eta^a \wedge \overline{\eta}^{\bar{a}}\equiv t\sum_a p^a w^a\wedge dy^a\, ,\\ 
        \Omega:=& \sqrt{\det(g_{a{\bar{b}}})} 
        \, \eta^1\wedge \eta^2\wedge \eta^3\equiv t\, e^{\phi_A}\, \eta^1\wedge \eta^2\wedge \eta^3\, .
    \end{align}
    We have normalized these in the usual way $\frac{1}{3!}J^3=\frac{i}{8}\Omega \wedge \overline{\Omega}=\text{Vol}(X_k)$. Under the O6-orientifold action we then have that
    \begin{equation}
        J\mapsto -J\, ,\quad \Omega\mapsto \overline{\Omega}\, .
    \end{equation}
    One computes 
    \begin{equation}
        dJ=t \sum_a p^a dw^a\wedge dy^a=-t \langle k,p \rangle d^3y =0\, ,
    \end{equation}
    and so \emph{$X_k$ is symplectic due to the PFV condition $\langle k,p \rangle=0$}. In particular, the torsion classes $W_1$, $W_3$ and $W_4$ vanish.

    With a bit of work one can further show that
    \begin{align}
         d\Omega=&\overline{W}_5 \wedge \Omega+W_2\wedge J+W_1\wedge \frac{1}{2}J\wedge J\, ,\\
         \overline{W_5}=& \bar{\del} A\, ,\quad W_2=-e^{\phi_A}K^{(8)}\, ,\quad W_1=e^{\phi_A}\frac{\langle k, p\rangle}{3\mathcal{V}_p t}\, ,
    \end{align}
    in terms of the real and \emph{primitive} (1,1)-form
    \begin{equation}
        K^{(8)}:=-\sum_a \hat{k}^a \frac{i}{2} r_{aa} \eta^a\wedge \overline{\eta^a}   -2i r^{a\bar{b}}\bar{\del}_{\bar{b}}A \epsilon_{abc}\eta^b\wedge \overline{\eta^c}\, ,\quad \hat{k}^a:=\kappa^{ab}k_b-\frac{\langle k,p\rangle }{6\mathcal{V}_p}p^a\, ,
    \end{equation}
    with $\kappa_{ab}:=\kappa_{abc}p^c$ and $\kappa^{ab}$ its inverse matrix. Again, the PFV condition $\langle k,p \rangle=0$ implies that the torsion class $W_1$ actually vanishes.

    The type IIA $B$-field background is given as
    \begin{equation}
        B=-b\sum_a p^a w^a\wedge dy^a\quad  \Rightarrow \quad  H_3=-b \, \langle k, p\rangle  d^3y=0\, ,
    \end{equation}
    and thus also the NSNS field strength vanishes as a consequence of the PFV condition $\langle k,p \rangle=0$.

    \subsubsection{RR sector}\label{subsubsec:toroidalPFVs:IIAsolution:RRsector}
    We now turn to the R-R field strengths. The RR 2-form receives a contribution from T-dualizing $F_3$,
    \begin{equation}
        F_2\supset F_2^{(A)}:=-\sum_a m^a w^a\wedge dy^a\quad \Rightarrow \quad dF^{(A)}_2=\langle \vec{m} , \vec{k}\rangle d^3y =-2Qd^3y\, ,
    \end{equation}
    which, in complex notation, can be written as
    \begin{equation}
        F_2^{(A)}= -\sum_a m^a \frac{i}{2} r_{aa}\eta^a\wedge \overline{\eta^a}\, .
    \end{equation}
    Another RR two-form component is obtained by T-dualizing $F_5$:
    \begin{equation}
        F_2\supset F_2^{(B)}:= \hat{*}d\chi \quad \Rightarrow \quad  dF_2^{(B)}= 2Qd^3y+j_{D6}(y;y_{D6})d^3y\, ,
    \end{equation}
    and in complex notation this component reads
    \begin{equation}
        F_2^{(B)}=-2i r^{a\bar{b}} \bar{\del}_{\bar{b}}A \epsilon_{acd} \eta^c \wedge \overline{\eta^d}+\left(i r^{a\bar{b}} \del_{\bar{b}}A \epsilon_{acd} \eta^c \wedge \eta^d+c.c.\right)\, .
    \end{equation}
    Putting both pieces together, $F_2\equiv F_2^{(A)}+F_2^{(B)}$, and decomposing by Hodge type, we get
    \begin{align}
        F_2|_{(1,1)}&\equiv F_2^{(8)}+J\wedge F_2^{(1)}\, ,\quad 
        F_2|_{(2,0)}\equiv  g^{a\bar{b}}\left(F_2^{(\bar{3})}\right)_{\bar{b}} \,\frac{1}{2}\Omega_{acd} \,\eta^c \wedge \eta^d\, ,\\
        F_2^{(8)}&=-\sum_a \hat{m}^a \frac{i}{2} r_{aa} \eta^a\wedge \overline{\eta^a}   -2i r^{a\bar{b}}\bar{\del}_{\bar{b}}A \epsilon_{abc}\eta^b\wedge \overline{\eta^c}\, ,\quad \hat{m}^a:=m^a-\frac{N_{bc}p^b p^c}{6\mathcal{V}_p}\\
         F_2^{(1)}&=-\frac{N_{ab}p^ap^b}{6t\mathcal{V}_p}\, ,\quad F_2^{(\overline{3})}=2i e^{-\phi_A}\bar{\del} A\, .
    \end{align}
    Now, due to the PFV conditions $\langle k,p\rangle= N_{ab}p^a p^c=0$ and $k_a=N_{ab}p^b\equiv \kappa_{ab} m^b$ various simplifications occur, and in particular we find that
    \begin{equation}
        F_2^{(1)}=0\, ,\quad e^{\phi_A}F_2^{(8)}=-W_2\, ,\quad e^{\phi_A}F_2^{(\overline{3})}=\overline{W}_5\, ,
    \end{equation}
    in terms of the torsion classes $W_2$ and $W_5$.
    
    Comparing with \cite{Grana:2004bg} we conclude that the RR-flux and the torsion classes are related precisely in the way needed to give consistent supersymmetric Minkowski vacua of type IIA string theory.\footnote{See second column of \textbf{IIA}-table of \cite{Grana:2004bg}, with convention $\beta=1$.}
    
    Furthermore, because the bulk components of $dF_2^{(A)}$ and $dF_2^{(B)}$ precisely cancel, the type IIA Bianchi identity is also satisfied
    \begin{equation}
        dF_2=j_{D6}(y;y_{D6})d^3 y\, ,
    \end{equation}
    with the expected localized D6-charge density
    \begin{equation}
        j_{D6}(y;y_{D6}):=\sum_{\alpha=1}^{2N}\delta^{(3)}(y-y_\alpha^{D6})-4\sum_{p=1}^{8}\delta^{(3)}(y-y_p^{O6})\, .
    \end{equation}
    This charge density is puzzling at first: we have 
    \begin{equation}
        \int_{T^3_y}j_{D6}(y;y_{D6})=-2Q\neq 0\, ,
    \end{equation}
    which looks like an uncanceled D6-brane tadpole! 
    
    More careful thought however reveals that this is not an inconsistency: $F_2$ is globally well-defined, but the integral over the $T^3_y$ coordinates is not a three-cycle in the twisted torus fibration, due to the monodromies \eqref{eq:monodromies}. Indeed its would-be Poincar\'e dual three-form $w^1\wedge w^2\wedge w^3$ is not closed.\footnote{See \cite{Andriolo:2018yrz} for a closely related discussion of this issue.}
    
    The fact that the base of the fibration is not a sub-manifold also implies that the generic fiber is at most a torsion element in homology \cite{Marchesano:2006ns}: for every closed three-form $\omega_3$ on the twisted six-torus, integration over the fiber indeed gives zero,
    \begin{equation}
        \int_{T^3_x,y=y_0}\omega_3 =\int \omega_3\wedge d^3y=\frac{1}{k_1}\int \omega_3 \wedge d(-dy^1\wedge w^1) = \int d\omega_3 \wedge dy^1\wedge w^1=0\, .
    \end{equation}
    In particular, the D6-branes as well as the O6-planes wrap torsional cycles \cite{Marchesano:2006ns,Andriolo:2018yrz}.\footnote{The homology groups of the nil-manifolds considered here can be found in Appendix B of \cite{Marchesano:2006ns}.} Their net-negative overall tension appears to be precisely what is needed to cancel the positive vacuum energy from the negative curvature of the twisted torus.
    
    Finally, one easily verifies that
    \begin{equation}
        F_4=B\wedge F_2\, ,\quad F_6=\frac{1}{2}B\wedge B\wedge F_2\, .
    \end{equation}
    We have therefore fully characterized the type IIA supergravity background. Its key features are
    \begin{enumerate}
        \item The geometric background $X_k$ is a smooth symplectic $SU(3)$-structure manifold.
        \item The only singular sources are D6-branes and O6-planes.
        \item The only non-vanishing RR-flux is $F_2$, which is primitive.
    \end{enumerate}

    \subsection{M-theory lift}\label{subsec:toroidalPFVs:MtheoryUplift}
    Any compactification with these properties should have an M-theory lift on a $7d$ spin-manifold $Y_7$, as a circle fibration over $X_k$: D6-branes are well known to correspond to local Taub-NUT geometries \cite{Sorkin:1983ns,Gross:1983hb}, O6-planes turn into Atiyah-Hitchin manifolds \cite{Seiberg:1996nz}, and $F_2$ is the Kaluza-Klein field strength that encodes the circle fibration \cite{Witten:1995ex}. Furthermore, due to unbroken supersymmetry in a Minkowski background the M-theory seven-fold must have holonomy group contained in $G_2$ \cite{Kaste:2003zd,Behrndt:2005im}. 
    
    Depending on how much SUSY is preserved, one should get 
    \begin{enumerate}
        \item $\mathcal{N}=4$: $Y_7=T^3\times K3$, and the holonomy group is $SU(2)$,
        \item $\mathcal{N}=2$: $Y_7=S^1\times CY_3$, with $CY_3$ a Calabi-Yau threefold. The holonomy group is $SU(3)$,
        \item $\mathcal{N}=1$: $Y_7$ is a $G_2$-manifold. I.e., the holonomy group is equal to $G_2$.
    \end{enumerate}
    To construct these manifolds explicitly, locally, we first write $F_2=dC_1$ in terms of a RR one-form potential $C_1$ on the twisted six-torus. Explicitly, one may express it as
    \begin{equation}
        C_1= -\sum_a \rho^a(x,y) dy_a\, ,\quad \rho^a(x,y)=-m^a x^a+\sum_{bc}  \frac{1}{2}\epsilon ^{abc} Q_by^b y^c+\delta\rho^a(y)\, ,
    \end{equation}
    with
    \begin{equation}
        (\vec{\del}_{y}\times \vec{\delta \rho})_a=\frac{\del_{y_a}\chi}{\hat{p}^a \xi^a}-Q_a y_a\, ,\quad Q_a:=-m^a k_a\quad  \text{(no sum over }a).
    \end{equation}
    Using this we may write down the M-theory lift of this solution family:
    \begin{equation}\label{eq:G2_metric}
        ds^2_M= \alpha^{-\frac{2}{3}} ds^2_{\mathbb{R}^{1,3}}
        +t^{\frac{2}{3}} \left[\mathcal{V}_p^{\frac{1}{3}}\left(\frac{(d\psi-C_1)^2}{\chi(y;y_{D6})}+ \chi(y;y_{D6}) d\hat{s}^2\right)+\sum_a\hat{p}^a\frac{(w^a)^2}{\xi^a}\right]\, ,
    \end{equation}
    with M-theory circle coordinate $\psi\simeq \psi+1$, and where the 11d spacetime is subject to the identification
    \begin{equation}
        (y^1,y^2,y^3,\psi)\sim (-y^1,-y^2,-y^3,-\psi) \, .
    \end{equation}
    Notably, in passing to 11d Einstein-frame, the metric has become a direct product between four dimensional Minkowski space and a Ricci-flat seven-fold $Y_7$.  
    
    By inspecting the metric \eqref{eq:G2_metric} we see that $Y_7$ is a $T^4$-fibration over the base $T^3_{\vec{y}}/\mathbb{Z}_2$,
    \begin{equation}\label{eq:G2_fibration_structure}
        T^4\hookrightarrow Y_7 \twoheadrightarrow T^3_{\vec{y}}/\mathbb{Z}_2\, ,
    \end{equation}
    with $T^4$ parameterized by $(x^1,x^2,x^3,\psi)$.
    
    The family of metrics \eqref{eq:G2_metric} is parameterized by 
    \begin{enumerate}
        \item the universal volume modulus $t$,
        \item $3N$ real moduli $y^{D6}_{a\alpha}$ parameterizing the $\delta$-sources in \eqref{eq:3d_electrostatic},
        \item two independent real parameters out of the $\vec{\xi}\in \mathbb{R}_+^3$, constrained to the surface $\xi^1\xi^2\xi^3=1$,
        \item one real parameter $c$ in the general solution \eqref{eq:chi_solution} to the 3d electro-static problem \eqref{eq:3d_electrostatic}, $\chi=\chi_0+c$.
    \end{enumerate}
    Each of these parameters is associated with a harmonic three-form, and is partnered with a massless axion from dimensional reduction of $C_3$. If the holonomy group is equal to $G_2$ we thus get\footnote{The amount of preserved SUSY depends on the choice of type IIB three form fluxes \cite{Frey:2002hf}. In \S\ref{subsec:toroidalPFVs:concrete_solutions} we will encounter concrete solutions preserving $4d$ $\mathcal{N}=4,2,1$ SUSY, corresponding to holonomy groups $SU(2)$, $SU(3)$ and $G_2$ respectively.}
    \begin{equation}
        b_3=b_4\geq 4+3N
    \end{equation}
    chiral multiplets, and additional contributions to the third and fourth Betti-numbers can (and will) come from non-universal moduli.
    
    For definiteness we specify four model independent four-forms as the Poincar\'e duals of the classes of the following three-cycles
    \begin{align}\label{eq:bulk_cycles}
        \Sigma_0:&\quad \left\{x^1=x^2=x^3\, ,\quad y_1=y_2=y_3\right\}\, ,\quad \text{Vol}(\Sigma_0)=t\sum_a p^a\, ,\\
        \Sigma_1:&\quad \left\{x^2=x^3=y_1=\psi=0\right\}\, ,\quad \text{Vol}(\Sigma_1)=t\mathcal{V}_p^{\frac{1}{3}} \xi^2 \xi^3 \int dy_2 dy_3 \,\chi(y_1=0,y_2,y_3)\, ,\\
        \Sigma_2:&\quad \left\{x^1=x^3=y_2=\psi=0\right\}\, ,\quad \text{Vol}(\Sigma_2)=t\mathcal{V}_p^{\frac{1}{3}} \xi^1 \xi^3 \int dy_1 dy_3 \,\chi(y_1,y_2=0,y_3)\, ,\\
        \Sigma_3:&\quad \left\{x^1=x^2=y_3=\psi=0\right\}\, ,\quad \text{Vol}(\Sigma_3)=t\mathcal{V}_p^{\frac{1}{3}} \xi^1 \xi^2 \int dy_1 dy_2 \,\chi(y_1,y_2,y_3=0)\, .
    \end{align}
    The associative three-form of the $G_2$ manifold is
    \begin{equation}
        \Phi:=e^{-\phi_A}\text{Re}(\Omega)-J\wedge (d\psi-C_1)\, ,
    \end{equation}
    and is indeed closed and co-closed. One easily checks that the volumes of the four three-cycles specified in \eqref{eq:bulk_cycles} are indeed the integrals over this form,
    \begin{equation}
        \text{Vol}(\Sigma_I)=\int_{\Sigma_I}\Phi\, .
    \end{equation}
    Na\"ively, the three-cycles $\Sigma_{1,2,3}$ come as three-parameter families parameterized by the values of the coordinates that are held fixed. For example, $\Sigma_1$ could be generalized to $y_1=\zeta^1$, $x^2=\zeta^2$, $x^3=\zeta^3$ in terms of three periodic parameters $\zeta^a$, and in this case $\text{Vol}(\Sigma_1)$ depends explicitly on $\zeta^1$. But in fact only $\zeta^{2,3}$ are allowed to continuously vary, while the monodromies \eqref{eq:monodromies} enforce $\zeta^1 k_1=0\mod 1$. Thus, the $\Sigma_{1,2,3}$ come in two-parameter families and the calibrated cycle volumes only depend on the homology class as they should.

    As in \cite{Sen:1997kz}, near any of the D6-brane locations the metric looks like a Taub-NUT solution, and, in particular, the 11d metric \eqref{eq:G2_metric} is smooth at $y=y^{D6}$.
    
    In addition to the chiral multiplets there are at least $N$ abelian vector multiplets. These are the M-theory lifts of the open string $U(1)$s that live on the D6-branes, and, as is well known, they correspond to the normalizable harmonic two-forms that arise in the local Taub-NUT geometries \cite{Ruback:1986ag}. In general, $U(1)$ gauge fields arise from the dimensional reduction of $C_3$, and thus we find second Betti number
    \begin{equation}
        b_2\geq N\, .
    \end{equation}
    
    In contrast to the D6-brane case, near an O6 plane the metric we have written down looks like a Taub-NUT solution with negative mass parameter, and suffers a local pathology \cite{Sen:1997kz,Hanany:2000fw}. We have 
    \begin{equation}
        \chi\approx const -\frac{1}{\pi r}\, ,
    \end{equation}
    with $r$ the radial distance from the O6-position, as measured with $d\hat{s}^2$. Thus for sufficiently small distance $r$ we get that $\chi$ turns negative and the solution breaks down. This problem is the type IIA/M-theory version of the singularity that arises near the location of localized sources of negative D3-charge in type IIB, as studied in \cite{Carta:2019rhx,Gao:2020xqh,Carta:2021lqg}. But, in the M-theory picture, the physical resolution of this pathology is completely geometric: the local singularity is simply replaced by gluing in Atiyah-Hitchin manifolds \cite{Seiberg:1996nz}. Thus, the metric we have written down in \eqref{eq:G2_metric} is only an approximation of the full $G_2$ metric. At generic points in $Y_7$ the approximation is valid when the constant term in \eqref{eq:chi_solution} is sufficiently large,
    \begin{equation}
        c\gg 1\, .
    \end{equation}
    Otherwise, using the language of \cite{Gao:2020xqh}, there is a ``singular bulk problem''. We emphasize here that if this problem is present, its resolution in the M-theory context is a purely classical geometric effect (see also \cite{Andriolo:2018yrz}). This is in contrast to the type II frames where this regime appears deeply quantum. We will return to this subject in \S \ref{sec:SingularBulk}.

    \subsection{Concrete solutions}\label{subsec:toroidalPFVs:concrete_solutions}
    Having discussed the general features of PFV flux vacua of $T^6/\mathbb{Z}_2$ and their type IIA and M-theory duals, we now turn to explicit solutions of the PFV constraints \S\ref{subsec:PFVs:PFVs}. We note that this has previously been studied in \cite{Cicoli:2022vny}, but our discussion will be substantially different. 
    
    First, existence of a flat direction implies the vanishing of the following determinant:
    \begin{equation}
        \det \mathtt{N}=0\, ,\quad \mathtt{N}:= \begin{pmatrix}
            0 & - k_b\\
            -k_a & \kappa_{abc}m^c
        \end{pmatrix}\overset{!}{=}0\, .
    \end{equation}
    For the intersection numbers of our diagonal $T^6$, this constraint can be written as a homogeneous quadratic Diophantine equation
    \begin{equation}\label{eq:DiophantineXYZ}
        X^2+Y^2+Z^2-2Z X-2Z Y-2 XY=0\, ,\quad (X,Y,Z):=\frac{1}{4}(-k^1m_1,-k^2m_2,-k^3m_3)\, ,
    \end{equation}
    where $(X,Y,Z)\in \mathbb{Z}^3$. D3-charge cancellation with fluxes and mobile D3 branes further implies that
    \begin{equation}\label{eq:D3_tadpoleXYZ}
        0\leq X+Y+Z\leq 8\, .
    \end{equation}
    It is easy to show that \eqref{eq:DiophantineXYZ} and \eqref{eq:D3_tadpoleXYZ} together imply that $X,Y,Z\geq 0$, and thus the solutions are a subset of the partitions of length three of $0,\ldots,8$. Up to permutations, these are given by $(X,Y,Z)=n(1,1,0)$ with $n=0,1,2,3,4$, as well as
    \begin{equation}
        (X,Y,Z)=(4,1,1)\, .
    \end{equation}
    For each of these the solutions for the flux quanta are immediately obtained, except for the case
    \begin{equation}
        0=Z=k^3m_3\, ,
    \end{equation}
    which has infinitely many solutions. But, by inspecting the resulting null vectors of $\mathtt{N}$ one sees that the only solutions with eigenvectors that point toward large complex structure and small string coupling are those with $k^3=m_3=0$. 
    
    The resulting 17 inequivalent configurations that satisfy all conditions are given in Table \ref{tab:Flux_solutions}. Out of these, only the trivial solution preserves $\mathcal{N}=4$ SUSY. In addition, there are 13 $\mathcal{N}=2$ solutions, one of which was studied in \cite{Kachru:2002sk}. Finally, there are three $\mathcal{N}=1$ solutions, including one that has previously been found in \cite{Cicoli:2022vny}.

    \begin{table}
        \centering
        \begin{tabular}{c|c|c|c|c|c|c|c}
            $\vec{m}$ & $\vec{k}$ & $\vec{p}$ & $N$ & SUSY & $U(1)$s & $\delta \tilde{J}$ & $ \delta(\tilde{\Omega},\tau)/(\tilde{J}\wedge \delta v)$\\[2pt]\hline 
            
             $0$ & $0$ & any & $16$ & $\mathcal{N}=4$ & $28$ & $9$ & $14$  \\ [5pt]
             
             $2(1,-1,0)$ & $2(-1,1,0)$ & $(\lambda,\lambda,1)$  & $12$ & $\mathcal{N}=2$ & $15$ & $4$ & $6$ \\[5pt]
            
             $2(2,-1,0)$ & $2(-1,2,0)$ & $(2\lambda,\lambda,1)$  & $8$ & $\mathcal{N}=2$ & $11$ & $4$ & $6$\\[2pt]
             $2(1,-1,0)$ & $2(-2,2,0)$ & $(\lambda,\lambda,2)$  & $8$ & $\mathcal{N}=2$ & $11$ & $4$ & $6$\\[2pt]
             $2(2,-2,0)$ & $2(-1,1,0)$ & $(\lambda,\lambda,1/2)$  & $8$ & $\mathcal{N}=2$ & $11$ & $4$ & $6$\\[5pt]
            
             $2(3,-1,0)$ & $2(-1,3,0)$ & $(3\lambda,\lambda,1)$  & $4$ & $\mathcal{N}=2$ & $7$ & $4$ & $6$\\[2pt]
             $2(1,-1,0)$ & $2(-3,3,0)$ & $(\lambda,\lambda,3)$  & $4$ & $\mathcal{N}=2$ & $7$ & $4$ & $6$\\[2pt]
             $2(3,-3,0)$ & $2(-1,1,0)$ & $(\lambda,\lambda,1/3)$  & $4$ & $\mathcal{N}=2$ & $7$ & $4$ & $6$\\[5pt]

             $2(4,-1,-1)$ & $2(-1,1,1)$ & $(1,1/2,1/2)$  & $4$ & $\mathcal{N}=1$ & $5$ & $4$ & $2$\\[2pt]
             $2(2,-1,-1)$ & $2(-2,1,1)$ & $(1,1,1)$  & $4$ & $\mathcal{N}=1$ & $5$ & $4$ & $2$\\[2pt]
             $2(1,-1,-1)$ & $2(-4,1,1)$ & $(1,2,2)$  & $4$ & $\mathcal{N}=1$ & $5$ & $4$ & $2$\\[5pt]

             $2(4,-1,0)$ & $2(-1,4,0)$ & $(4\lambda,\lambda,1)$  & $0$ & $\mathcal{N}=2$ & $3$ & $4$ & $6$\\[2pt]
             $2(1,-1,0)$ & $2(-4,4,0)$ & $(\lambda,\lambda,4)$  & $0$ & $\mathcal{N}=2$ & $3$ & $4$ & $6$\\[2pt]
             $2(4,-4,0)$ & $2(-1,1,0)$ & $(\lambda,\lambda,1/4)$  & $0$ & $\mathcal{N}=2$ & $3$ & $4$ & $6$\\[2pt]

            $2(2,-2,0)$ & $2(-2,2,0)$ & $(\lambda,\lambda,1)$  & $0$ & $\mathcal{N}=2$ & $3$ & $4$ & $6$\\[2pt]
             $2(4,-2,0)$ & $2(-1,2,0)$ & $(\lambda,\lambda/2,1/2)$  & $0$ & $\mathcal{N}=2$ & $3$ & $4$ & $6$\\[2pt]
             $2(2,-1,0)$ & $2(-2,4,0)$ & $(2\lambda,\lambda,2)$  & $0$ & $\mathcal{N}=2$ & $3$ & $4$ & $6$\\[2pt]
        \end{tabular}
        \caption{All inequivalent PFVs on $T^6/\mathbb{Z}_2$ for diagonal $T^6=T^2\times T^2\times T^2$. The $(\vec{m},\vec{k})$ are the RR and NSNS fluxes, $\vec{p}$ is the flat direction in complex structure moduli space, $N$ is the number of mobile D3-branes, and we indicate the amount of unbroken SUSY, the rank of the gauge group, and the number of massless real K\"ahler moduli ($\delta \tilde{J}$), and combinations of complex structure moduli and axio-dilaton modulo redundant ones ($\delta (\tilde{\Omega},\tau)/(\tilde{J}\wedge \delta v)$). All solutions with extended SUSY have an additional diagonal flat direction that we parameterize by $\lambda\in \mathbb{C}$.}
        \label{tab:Flux_solutions}
    \end{table}
    
    \subsubsection{$\mathcal{N}=4$ solution}\label{subsubsec:toroidalPFVs:concrete_solutions:N=4}
    The most obvious and of course well known solution is the $\mathcal{N}=4$ case, where all fluxes vanish. The massless degrees of freedom are the gravity multiplet, and $22$ vector multiplets, and the generic gauge group is $U(1)^{28}$. 
    
    The type IIA manifold is a $T^6$, and the M-theory uplift is $T^3\times$K3, with K3-metric
    \begin{equation}\label{eq:K3_metric}
        ds^2_{K3}=t^{\frac{2}{3}}\mathcal{V}_p^{\frac{1}{3}}\left(\frac{(d\psi-C_1)^2}{\chi(y;y_{D6})}+ \chi(y;y_{D6}) \sum_a \hat{p}^a\xi^a (dy_a)^2\right)\, .
    \end{equation}
    In general, this metric is only an approximation of the full K3-metric. But, as we shall see momentarily, the approximation is exponentially good when the parameter $c$ in \eqref{eq:chi_solution} is large.
    
    First, when all D6-branes are coincident with $O6$-planes we get $\chi=const$ and we recover the orbifold limit $K3=T^4/\mathbb{Z}_2$. In this case the metric \eqref{eq:K3_metric} becomes exact. Moving away from the orbifold limit in the D6-brane position moduli space amounts to resolving the 16 local $A_1$ singularities with independent $\mathbb{P}^1$s, i.e., gluing in 16 copies of local Eguchi-Hanson metrics, and in general this is not captured by the metric \ref{eq:K3_metric}.

    However, when $c\gg 1$ in \eqref{eq:chi_solution}, the size of the $S^1$ parameterized by $\psi$ is much smaller than the $\vec{y}$-directions. Then, close to the orbifold limit, the 16 Eguchi-Hanson spaces come in eight pairs, where each pair is a fixed point on the base $T^3/\mathbb{Z}_2$ over which lie two Eguchi-Hanson spaces at opposite ends of the $\psi$-interval $S^1/\mathbb{Z}_2$. Stripping away the mobile D6-branes even to generic positions leaves behind an Atiyah-Hitchin manifold \cite{Atiyah:1985fd}, which at long distances is exponentially well-approximated by a Taub-NUT solution with negative mass parameter \cite{Sen:1997kz}. This is precisely the approximation implicit in \ref{eq:K3_metric}, which can be viewed as a compact version of the interpolation between Taub-NUT and Atiyah-Hitchin manifolds described in \cite{Sen:1997kz}. A distance $r$ away from the O6-plane positions --- as measured with $d\hat{s}^2$ --- the full K3 metric is equal to \eqref{eq:K3_metric} up to corrections of order $e^{-2cr}$ \cite{Hanany:2000fw}, and only within a radius $\sim 1/c$ of the O6-positions does the metric receive $\mathcal{O}(1)$ corrections.\footnote{The absence of perturbative corrections to \eqref{eq:K3_metric} in $1/c$ appears to be the M-theory dual of the vanishing of perturbative warping corrections to the K\"ahler potential, as found for toroidal orientifolds in \cite{Frey:2013bha}. I am thankful to Andrew Frey for an interesting discussion about this.}
    
    Thus, to conclude, the metric \eqref{eq:K3_metric} is a good approximation at large $c$, and is the M-theory dual of the weak warping approximation in the type II descriptions.
    
    \subsubsection{$\mathcal{N}=2$ solutions}\label{subsubsec:toroidalPFVs:concrete_solutions:N=2}
    As indicated Table \ref{tab:Flux_solutions}, there are $13$ flux vacua that preserve $\mathcal{N}=2$ supersymmetry. These solutions are all very similar, essentially only differing in the number of mobile $D3$-branes, and so we will restrict attention to a single example,
    \begin{equation}
        \vec{m}=-\vec{k}=2(2,-2,0)\, ,
    \end{equation}
    in which case $N=0$.
    
    We will begin by discussing the 4d effective theory. First, after imposing primitivity of the fluxes a four-dimensional K\"ahler moduli space remains, parameterized by the three diagonal moduli $\tilde{t}^a$, $a=1,2,3$, plus an off-diagonal perturbation of the K\"ahler form
    \begin{equation}
        \delta \tilde{J} = g_{1\bar{2}}\frac{i}{2}d\tilde{\phi}^1\wedge \overline{d\tilde{\phi}^2}+c.c.\, ,\quad g_{1\bar{2}}\in i \mathbb{R}\, .
    \end{equation}
    Three combinations $\mathcal{A}_\mu^{1,2,3}$ of the 12 abelian vector fields remain massless, and correspond to
    \begin{equation}
        C_2+\tau \tilde{B}_2=i\mathcal{A}^1\wedge d\tilde{\phi}^3+\left(\mathcal{A}^2+i \mathcal{A}^3\right)\wedge d\overline{\tilde{\phi}^3}\, ,
    \end{equation}
    and thus the generic gauge group is $U(1)^3$.
    One combination of vector fields is part of the gravity multiplet, and two others are the spin one components of two vector multiplets. 
    
    As three $U(1)$s remain unbroken, there are also three additional $C_4$-axions beyond the universal ones, for a total of six:
    \begin{align}
        C_4=&c_1\, d\tilde{x}^2\wedge dy_2\wedge d\tilde{x}^3\wedge dy_3+c_2\, d\tilde{x}^1\wedge dy_1\wedge d\tilde{x}^3\wedge dy_3+c_3 d\tilde{x}^1\wedge dy_1\wedge d\tilde{x}^2\wedge dy_2\nonumber\\
        &+\left(c_4 \,d\tilde{x}^1\wedge d\tilde{x}^2+ c_5 \,dy_1\wedge dy_2+c_6 \,(dy_1\wedge d\tilde{x}^2+d\tilde{x}^1\wedge dy_2)\right)\wedge d\tilde{x}^3\wedge dy_3\, .
    \end{align}

    Turning to the complex structure and dilaton sector, by construction, the flux superpotential is F-flat along the family of PFVs
    \begin{equation}
        z^a\equiv z^{aa}=p^a \tau\, ,\quad p^a=(\lambda,\lambda,1)\, ,\quad \lambda\in \mathbb{C}\, ,
    \end{equation}
    but two additional non-diagonal deformations $z^{12}$ and $z^{21}$ also remain unconstrained. One linear combination of these latter two is in the image of $J\wedge \bullet $, and therefore we get three complex moduli in the $(\tau,z^{ab})$ sector.

    The fact that this solution preserves $\mathcal{N}=2$ SUSY is seen easily as follows \cite{Frey:2002hf}. For simplicity we set $\tau=i$ and $\lambda=1$, in which case the complex three form field strength takes the form
    \begin{equation}
        G_3=F_3-\tau H_3=i \left(\overline{\tilde{\phi}^1}\wedge \tilde{\phi}^2 - \tilde{\phi}^1 \wedge \overline{\tilde{\phi}^2}\right) \wedge \tilde{\phi}^3\, .
    \end{equation}
    Evidently, for our choice of complex structure, it is of Hodge type $(2,1)$. But it is also of Hodge type $(2,1)$ if we redefine $(\tilde{\phi}^1,\tilde{\phi}^2) \mapsto (\overline{\tilde{\phi}^1},\overline{\tilde{\phi}^2})$, and thus the flux preserves $4d$ $\mathcal{N}=2$.

    Two complex combinations of the $16$ real closed-string moduli become the scalar components of the two vector multiplets, and the remaining $12$ real moduli therefore reside in $3$ closed-string hypermultiplets.
    
    All the $\mathcal{N}=2$ solutions in Table \ref{tab:Flux_solutions} have the above properties, except that some also have mobile D3 branes $N\in \{0,2,4,6,8,12\}$. Each mobile D3-brane contributes a vector and a hypermultiplet, and thus the total number of vector and hypermultiplets in this class of $\mathcal{N}=2$ flux vacua is
    \begin{equation}
        n_V=2+N\, ,\quad n_H=3+N\, .
    \end{equation}
    
    We now turn to the ten and eleven dimensional solutions. Nothing in addition needs to be said about the type IIB solution, but the type IIA solution has the important feature that the monodromies \eqref{eq:monodromies} reduce to
    \begin{align}
        I_1:&\, (y_2,x^1)\simeq (y_2+1,x^1-4 y_3)\, ,\\
        I_2:&\, (y_3,x^2)\simeq (y_3+1,x^2+4 y^1)\, ,
    \end{align}
    and thus the twisted six-torus becomes a warped product between a twisted five-torus and a circle parameterized by the $x^3$ coordinate. Furthermore, as $m^3=0$, the RR 1-form is also independent of $x^3$. Therefore, as it must, the M-theory metric \eqref{eq:G2_metric} reduces to a direct product between an $S^1$ parameterized by $x^3$ and a Calabi-Yau threefold with metric
    \begin{equation}
        ds^2_{CY_3}= (t\lambda)^{\frac{2}{3}}\left(\frac{(d\psi-C_1)^2}{\chi(y;y_{D6})}+ \chi(y;y_{D6}) d\hat{s}^2\right)+\lambda\sum_{a=1}^2\frac{(w^a)^2}{\xi^a}\, ,
    \end{equation}
    where for simplicity we have taken $\lambda\in \mathbb{R}_+$.
    
    As in the $\mathcal{N}=4$ case this metric approximates the Calabi-Yau metric exponentially well at large $c$, and makes manifest an SYZ-fibration over a base $T^3/\mathbb{Z}_2$ with fiber-$T^3$ parameterized by $(x^1,x^2,\psi)$. The family of metrics admits both LCS and Large Volume limits:
    \begin{align}
        \text{LCS:}\quad &t=c=\frac{1}{\lambda}\longrightarrow \infty\, ,\\
        \text{Large Volume:}\quad & \lambda=t^2\longrightarrow \infty\, .
    \end{align}
    
    Finally, given the fact that M-theory on $CY_3\times S^1$ leads to $h^{1,1}$ vector multiplets and $h^{2,1}+1$ hypermultiplets, we conclude that the Calabi-Yau threefolds constructed in this way have Hodge numbers
    \begin{equation}
        h^{1,1}=h^{2,1}=N+2\, .
    \end{equation}
    It would be interesting to match these Calabi-Yau threefolds with known algebraic constructions, such as those listed in \cite{Candelas:2016fdy}.

    \subsubsection{$\mathcal{N}=1$ solutions}\label{subsubsec:toroidalPFVs:concrete_solutions:N=1}
    Finally, we turn to solutions with minimal supersymmetry in four dimensions. As a concrete example we will discuss the flux choice
    \begin{equation}
        \vec{m}=-\vec{k}=2(2,-1,-1)\, ,
    \end{equation}
    which has also previously been discussed in \cite{Cicoli:2022vny}.

    One linear combination $\mathcal{A}$ of the closed string U(1) gauge fields remains massless,
    \begin{equation}\label{eq:bulkU1_N=1example}
        C_2+\tau \tilde{B}_2=\mathcal{A}\wedge d\tilde{\phi}^1\, ,
    \end{equation}
    and correspondingly, there is one additional massless $C_4$ axion beyond the three universal ones,
    \begin{align}
        C_4=&c_1\, d\tilde{x}^2\wedge dy_2\wedge d\tilde{x}^3\wedge dy_3+c_2\, d\tilde{x}^1\wedge dy_1\wedge d\tilde{x}^3\wedge dy_3+c_3 d\tilde{x}^1\wedge dy_1\wedge d\tilde{x}^2\wedge dy_2\nonumber\\
        &+c_4\left(d\tilde{x}^2\wedge dy_3-dy_2\wedge d\tilde{x}^3 \right)\, .
    \end{align}
    These four $C_4$ axions pair into complex scalars with four massless K\"ahler moduli, corresponding to the most general K\"ahler class that leaves the fluxes primitive:
    \begin{equation}
        \widetilde{J}
        =\sum_{a=1}^3 \tilde{t}^a d\tilde{x}^a\wedge dy^a+\tilde{t}^4\left(dx^2\wedge dy^3-dy^2\wedge dx^3\right)\, .
    \end{equation}
    
     It is straightforward to show that, for $\tau=i$, the complex three-form flux $G_3:= F_3-\tau H_3$ can be written as
    \begin{equation}
        G_3=i\left(d\tilde{\phi}^1\wedge (-d\tilde{\phi}^2 \wedge d\overline{\tilde{\phi}}^3+
         d\tilde{\phi}^3 \wedge d\overline{\tilde{\phi}}^2)+2d\tilde{\phi}^2\wedge d\tilde{\phi}^3 \wedge d\overline{\tilde{\phi}}^1\right)\, ,
    \end{equation}
    and is thus indeed of Hodge type (2,1). As it is \emph{not} of Hodge type (2,1) with respect to any of the other inequivalent choices of complex structures on the $T^6$ this solution indeed preserves $\mathcal{N}=1$ supersymmetry in four dimensions. 
    
    All deformations of the complex structure away from the PFV locus are rendered massive, and we thus end up with a moduli space of $\mathcal{N}=1$ vacua parameterized by a single chiral multiplet from the complex structure and dilaton sector, plus four chiral multiplets from the K\"ahler moduli and $C_4$ axions, and $3N=12$ chiral multiplets from open strings. In addition, we get $N=4$ abelian vector multiplets from the open-string U(1)s and another one from \eqref{eq:bulkU1_N=1example}.

    Thus, the M-theory lift of the type IIA mirror duals yields three $T^4$-fibered $G_2$-manifolds, all of which have the same non-trivial Betti numbers
    \begin{equation}
        b_2=b_5=5\, ,\quad b_3=b_4=17\, .
    \end{equation}

    \section{The general Calabi-Yau case}
    \label{sec:GeneralCYcase}

    In this section we would like to extend our discussion of the relation between perturbatively flat vacua of type IIB string theory, the mirror dual type IIA compactifications and their M-theory lifts beyond the toroidal case of the previous section \ref{sec:toroidalPFVs}. We will argue that PFVs are a concrete class of flux vacua that generally have dual descriptions as both $SU(3)$ structure backgrounds with RR two-form flux in type IIA string theory, and $G_2$-holonomy compactifications of M-theory, as in \cite{Gukov:2002jv}, and as studied from the perspective of generalized complex geometry in \cite{Grana:2004bg,Grana:2005sn}.

    \subsection{Flux quantization in type IIA string theory}\label{subsec:GeneralCYcase:IIA_flux_quantization}
	We begin this discussion by recalling the Dirac quantization conditions for the RR field strengths in type IIA string theory, as these are slightly non-standard. 
	
	In ten dimensions we have the RR $p$-form field strengths $F_p$ with $p$ even, which, away from sources, satisfy non-standard Bianchi identities
	\begin{equation}\label{eq:RR_Bianchi_sourceless}
		0=d F_p+H\wedge F_{p-2}\, ,
	\end{equation}
	where $H=dB$ is the NSNS field strength (see for example \cite{Grana:2000jj}). Here, we are working in the ``democratic formulation'' where $p\in \{0,2,4,6,10\}$ and impose the self duality condition $*F_{p}=(-1)^{\frac{p}{2}}F_{10-p}$. 
	
	It is useful to consider the formal sum $F:=\sum_p F_p$ over all the RR field strengths, and define $\check{F}:=\sum_{p}(-1)^{\text{int}[\frac{p}{2}]}F_p$ for any polyform $F$.
	In terms of these we may write the Bianchi identities/equations of motion and self-duality constraints collectively as
	\begin{equation}
		\mathcal{D_+}F=0\, ,\quad \mathcal{D}_- \check{F}=0\, ,\quad  *F=\check{F}\, ,
	\end{equation}
	with covariant derivative $\mathcal{D}_{\pm}:=d\pm H_3\wedge$.
	
	This makes it clear that away from charged sources we can write the field strengths in terms of the conventionally defined RR $p-1$ form potentials $C_{p-1}$ as
	\begin{equation}
		F=\mathcal{D}_+C\, ,\quad C=\sum_{n=0}^5 C_{2n-1}\, .
	\end{equation}
	
	In the presence of RR-charges,
	$S\supset 2\pi\int C\wedge \check{J}_{RR}$, the Bianchi identities and equations of motion get modified to
	\begin{equation}
		\mathcal{D}_+F=J_{RR}\, ,\quad \mathcal{D}_-\check{F}=\check{J}_{RR}\, ,
	\end{equation}
	where $J_{RR}$ is an odd-degree polyform.
	
	The RR charges are D-branes wrapped on cycles $\Sigma$ with gauge bundles $F_\Sigma$ \cite{Green:1996dd,Cheung:1997az,Minasian:1997mm}
	\begin{equation}
		S_{CS}= 2\pi \int_{\Sigma} C\wedge e^{\frac{F_{\Sigma}}{2\pi}+B} \wedge \sqrt{\frac{\hat{A}(T\Sigma)}{\hat{A}(N\Sigma)}}\, ,
	\end{equation}
	where $\hat{A}$ denotes the A-roof genus, and $T\Sigma$ and $N\Sigma$ are the tangent respectively normal bundle of $\Sigma$. Thus, for a single such brane source we get
	\begin{equation}\label{eq:RR_Bianchi_Dbrane_sources}
		\check{J}_{RR}=e^{\frac{F_{\Sigma}}{2\pi}+B} \wedge \sqrt{\frac{\hat{A}(T\Sigma)}{\hat{A}(N\Sigma)}}\wedge \delta(\Sigma)\equiv e^{B}\wedge Q_{(\Sigma,F_{\Sigma})}\quad \Rightarrow\quad J_{RR}=e^{-B}\wedge \check{Q}_{(\Sigma,F_{\Sigma})}\, .
	\end{equation}
	We now consider type IIA string theory on a Calabi-Yau threefold $X$. Then all possible RR fluxes on internal cycles can be generated by crossing sufficiently general D8/D6/D4/D2 domain walls in the four extended directions of spacetime. Decomposing the RR field strength as
	\begin{equation}
		F=F_{\text{int}}+d^4x\wedge \hat{F}_{\text{int}}\, ,
	\end{equation}
	with $F_{\text{int}}$ and $\hat{F}_{\text{int}}$ polyforms defined on the internal space $X$, for any $F_{\text{int}}$ the self-duality condition is solved by setting $ \hat{F}_{\text{int}}\sim *_6 F_{\text{int}}$. Crossing a D-brane domain wall of charge $Q_{(\Sigma,F_{\Sigma})}=Q^{\text{int}}_{(\Sigma,F_{\Sigma})}\wedge \delta(z)$, with $z$ some spatial coordinate of $\mathbb{R}^{1,3}$, turns on RR-flux
	\begin{equation}
		\mathbb{F}:=e^{-B} \wedge F_{\text{int}}= \check{Q}^{\text{int}}_{(\Sigma,F_{\Sigma})}\, ,
	\end{equation}
	which is closed and single valued under the monodromies of the $B$-field.
	
	This in turn generates a superpotential in the four dimensional effective theory \cite{Grimm:2004ua}
	\begin{equation}\label{eq:RR_superpotential_IIA_formal}
		W^{RR}_0(z)=\int_X e^{B+iJ}\wedge \check{Q}^{\text{int}}_{(\Sigma,F_{\Sigma})}=\int_X e^{B_2+iJ}\wedge \mathbb{F}\, .
	\end{equation}
	The integrals of the collective RR field strength $\mathbb{F}$ over closed submanifolds  in $X$ are well-defined flux numbers that fully characterize it. For a suitably chosen basis of even-dimensional cycles of $X$, the corresponding flux numbers are identified with the ones we defined earlier:
	\begin{equation}
		m^0\equiv \int_X \mathbb{F}\wedge [\text{pt.}]\, ,\quad m^a\equiv \int_X \mathbb{F}\wedge [\Sigma_2]^a\, ,\quad e_a\equiv \int_X \mathbb{F}\wedge [\Sigma_4]_a\, ,\quad e_0\equiv \int_X \mathbb{F}\, ,
	\end{equation}
	where $[\text{pt.}]\in H^6(X,\mathbb{Z})$ is Poincar\'e dual to a point in $X$, $[\Sigma_2]^a$ is a basis of $H^4(X,\mathbb{Z})$, and $[\Sigma_4]_a$ is the dual basis of $H^2(X,\mathbb{Z})$. 
	
	Indeed, in terms of this data the RR superpotential \eqref{eq:RR_superpotential_IIA_formal} matches the expression \eqref{eq:RR_superpotential} under the identification
	\begin{equation}
		z^a\equiv \int_X  J_{\mathbb{C}}\wedge [\Sigma_2]^a\, ,\quad J_{\mathbb{C}}:=B_2+iJ\, ,
	\end{equation}
	and with $\kappa_{abc}:=\int_X [\Sigma_4]_a\wedge [\Sigma_4]_b\wedge [\Sigma_4]_c$ the triple intersection form on $X$.
	
	This form of the superpotential is of course very well known, and widely used in attempts to stabilize moduli in Calabi-Yau compactifications of type IIA string theory (starting with \cite{DeWolfe:2005uu}). However, a point less frequently emphasized is that the flux numbers $(m^0,m^a,e_a,e_0)$ that appear in $W^{RR}$ do not in general satisfy integer quantization conditions. Rather, one must derive the proper flux quantization conditions from a basis of the space of allowed forms $Q_{(\Sigma,F_{\Sigma})}$ (i.e., a K-theory basis). One can use a basis of branes wrapped on even-dimensional cycles with choices of gauge bundles such that Freed-Witten anomalies are canceled, and arrives at the shifted quantization conditions
	\begin{equation}\label{eq:IIA_flux_quantization}
		m^0,m^a\in \mathbb{Z}\, ,\quad \tilde{e}_a\equiv e_a-\frac{1}{2}\mathbb{A}_{ab}m^b+\frac{c_a}{24}m^0\in \mathbb{Z}\, ,\quad \tilde{e}_0\equiv e_0+\frac{c_a}{24}m^a\in \mathbb{Z}\, .
	\end{equation}
	Here, $c_a:=\int_X c_2(TX)\wedge [\Sigma_4]_a$ and the symmetric tensor $\mathbb{A}:\, H^2(X,\mathbb{Z}_2)\times H^2(X,\mathbb{Z}_2)\rightarrow \mathbb{Z}_2$ is defined in components as
	\begin{equation}
		\mathbb{A}_{ab}:=\begin{cases}
			\kappa_{aab}& a\leq b\\
			\kappa_{abb}& \text{else}
		\end{cases}\mod 2\, .
	\end{equation}
	Of course, even though the RR-flux $\mathbb{F}$ does not take values in even-dimensional de Rham cohomology, the Dirac quantization conditions are met: upon lassoing a D-brane probe around a cycle $\Sigma'$ its partition function (in euclidean signature) picks up a phase
	\begin{equation}
		\text{exp}\left(2\pi i \left\langle  Q^{\text{int}}_{(\Sigma,F_\Sigma)}, Q^{\text{int}}_{(\Sigma',F_{\Sigma'})} \right\rangle \right)\, ,
	\end{equation}
	where $\langle Q,Q'\rangle:=\int_X \check{Q}\wedge Q'$ is the anti-symmetric K-theory intersection pairing. As the latter is integral, there is no anomaly.
	
	In conclusion, we have now fully specified the data that goes into the prepotential \eqref{eq:prepotential}, and matched the integral fluxes of the RR three-form in type IIB string theory with suitably shifted fluxes of the RR forms in type IIA string theory. Importantly, the conventionally defined RR field strengths in type IIA string theory satisfy in general non-integral quantization conditions.  
	
	Finally, as is well known, the match between the RR superpotential in both formulations allows one to compute the entire series of worldsheet instanton corrections to the type IIA flux superpotential in terms of the periods of $\widetilde{X}$.\footnote{More precisely, the correction term $\delta W^{RR}$ in \eqref{eq:RR_superpotential} also contains a perturbative $\alpha'^3$ correction}
	
	\subsection{Mirror dual of RR-fluxes in PFVs}\label{subsec:GeneralCYcase:RRfluxes}
	Having understood the mirror correspondence of RR fluxes, we are ready to state the meaning of the special ansatz in fluxes
	defined around \eqref{eq:integrality_conditions}. To facilitate the discussion, for now, we set to zero the $H_3$ flux in type IIB and consider only the ``parent'' Calabi-Yau compactification without the orientifold involution. Even though NSNS fluxes are turned off for now, we will still assume that the relations \eqref{eq:PFV_locus} hold, but without assigning any independent physical meaning to the $k_a$ that appear in \eqref{eq:PFV_locus}.
	
	The type IIA interpretation of our choice of RR fluxes is rather simple: the integrality conditions \eqref{eq:integrality_conditions} ensure that one can consistently set the internal four and six-form flux to zero,
	\begin{equation}
		\mathbb{F}|_{\text{4-form}}= \mathbb{F}|_{\text{6-form}}=0\, ,
	\end{equation}
	and the further restriction $m^0=0$ becomes the vanishing of the Romans mass,
	\begin{equation}
		F_0\equiv \mathbb{F} |_{\text{0-form}}=0\, .
	\end{equation}
	Thus, the physical RR field strengths are given by
	\begin{equation}\label{eq:PFV_IIA_fluxes}
		F_0= 0\, ,\quad 
		F_2= m^a[\Sigma_4]_a\, ,\quad 
		F_4= B_2\wedge F_2\, ,\quad 
		F_6\simeq \frac{1}{2}B_2\wedge B_2\wedge F_2\, ,
	\end{equation}
	where $\simeq$ denotes retaining only the internal components.
	
	Using the conditions \eqref{eq:PFV_Diophantine_eq} and \eqref{eq:PFV_locus} 
    \begin{equation}\label{eq:F2_primitivity}
        \int_X F_2\wedge J\wedge J= \kappa_{abc}m^a p^b p^c= N_{ab}p^a p^b=\langle k,\vec{p}\rangle = 0\, ,
    \end{equation}
    and thus the RR two-form flux is \emph{primitive}.

    \subsection{Mirror dual of NSNS fluxes in PFVs}\label{subsec:GeneralCYcase:NSNSfluxes}
	Next, we turn to the type IIA interpretation of non-vanishing $H_3$ flux in the type IIB picture. For simplicity we will first discuss this again in the absence of an orientifold projection and also for vanishing RR flux. The discussion is fairly analogous to the construction of the twisted torus in \S \ref{sec:toroidalPFVs}.
	
	Of course, turning on a generic $\widetilde{H}_3$ flux in the type IIB solution does not have a geometric description in type IIA string theory. However, our special ansatz \eqref{eq:PFV_Hflux} amounts to a field strength $\widetilde{H}_3$ with precisely one leg along the SYZ fiber, and falls into the subclass of \emph{purely electric} NSNS fluxes whose role in the mirror correspondence was studied in detail in \cite{Kachru:2002sk,Gurrieri:2002wz}, whose discussion we will follow extensively in this section. Importantly, purely electric fluxes dualize to certain globally geometric --- though no longer Ricci flat --- $SU(3)$ structure compactifications of type IIA string theory, as in the simpler toroidal setup we discussed in \S\ref{subsec:toroidalPFVs:IIAsolution}.
	
	In the LCS limit one can describe the metric of the type IIA mirror threefold, which we will again denote by $X_k$, somewhat explicitly: one again begins by writing the type IIB B-field as
	\begin{equation}
		\widetilde{B}_2=\sum_{\alpha=1}^3 A^\alpha\wedge d\tilde{x}_\alpha\, ,
	\end{equation}
	with $\tilde{x}^\alpha\simeq \tilde{x}^\alpha+1$ the three periodic coordinates on the $T^3$ fiber of the SYZ fibration of $\tilde{X}$, and $A^\alpha$ a triplet of one-form gauge potentials on the three-dimensional base $\mathcal{B}$ of the SYZ fibration. 
	
	In the LCS limit the Calabi-Yau threefold $X$ is fibered by a flat three-torus, and thus its metric can be written as
	\begin{equation}\label{eq:CY_metric_LCS}
		ds^2=ds^2_{\mathcal{B}}+h_{\alpha\beta}(y)w^\alpha\otimes w^\beta\, ,
	\end{equation}
    where $ds^2_{\mathcal{B}}$ is the metric on the SYZ-base, parameterized by coordinates $y_{\alpha}$, and
    \begin{equation}
        w^\alpha= dx^\alpha+\tilde{A}^{\alpha}\, ,\quad \tilde{A}^\alpha:=\tilde{A}^{\alpha\beta}(y) dy_\beta\, ,
    \end{equation}
    in terms of a triplet of one-forms $\tilde{A}^{\alpha\beta}dy_\beta$, and $T^3$ coordinates $x^\alpha\simeq x^\alpha+1$. The geometric twist induced by the non-vanishing NSNS flux on the type IIB side then leads to the shift
    \begin{equation}\label{eq:Htwist_substitution}
        w^{\alpha}\longrightarrow w^{\alpha}+A^\alpha\, .
    \end{equation}
	If the triplet of Kaluza-Klein connections $A^\alpha$ are flat then $X_k\equiv X$ is simply the mirror dual Calabi-Yau threefold, but if $H_3\neq 0$ the $A^\alpha$ carry field strength, amounting to a non-trivial twist of the $T^3$ fibration over the base. The resulting manifold $X_k$ then only has $SU(3)$ \emph{structure} rather than $SU(3)$ holonomy.
	
	Given any closed $p$-form $\omega_p$ on $X$, there exists a corresponding twisted form $\omega_p^{(k)}$ on $X_k$ that is constructed by first expanding it in powers of $dy_\alpha$ and $w^\alpha$, and then substituting \eqref{eq:Htwist_substitution}.
	
	We will denote the twisted versions of the closed $p$-forms on $X$ by
	\begin{equation}
		([\Sigma_3]_I,[\Sigma_3]^I)\overset{\text{twist}}{\longrightarrow} (\alpha_I,\beta^I)\, ,\quad ([\Sigma_4]_a,[\Sigma_2]^a)\overset{\text{twist}}{\longrightarrow} (\omega_a,\omega^a)\, .
	\end{equation}
	The globally defined holomorphic three-form $\Omega^{(k)}$ and complexified $(1,1)$ form $J_{\mathbb{C}}=iJ+B_2$ on $X_k$ are obtained from those of $X$ by replacing all closed forms by their twisted cousins:
	\begin{align}
		J_{\mathbb{C}}=z^a [\Sigma_4]_a\quad &\longrightarrow \quad z^a \omega_a\, ,\\
		\Omega=\tilde{Z}^I [\Sigma_3]_I-\tilde{\mathcal{F}}_I [\Sigma_3]^I \quad &\longrightarrow \quad \tilde{Z}^I \alpha_I-\tilde{\mathcal{F}}_I \beta^I\, .
	\end{align}
	While the twisted forms satisfy the same intersection properties as their untwisted cousins, they are in general not closed:
	\begin{equation}
		d\omega_a=k_a \beta^0\, ,\quad d\omega^a=0\, ,\quad  d\alpha_0=K:=k_a\omega^a\, , \quad d\alpha_i=d\beta^I=0\, ,
	\end{equation}
	and this implies that neither $\Omega$ nor $J_{\mathbb{C}}$ are generally closed. In the standard normalization of $\Omega$ one finds
	\begin{align}\label{eq:dOmega_IIA}
		d\Omega&=W_2\wedge J+W_1\wedge \frac{1}{2}J \wedge J\, , \\
        W_2&=-e^{\phi_A}J\wedge K^{(8)}\, ,\quad W_1= e^{\phi_A}\frac{\langle \vec{k},\vec{p}\rangle}{3\mathcal{V}_p t}\, ,
	\end{align}
    with 
    \begin{equation}
        K^{(8)}=-\sum \hat{k}^a[\Sigma_4]_a\, ,\quad \hat{k}^a=\kappa^{ab}k_b-\frac{\langle \vec{k},\vec{p}\rangle}{6\mathcal{V}_p}p^a\, ,
    \end{equation}
    largely echoing \S\ref{subsec:toroidalPFVs:IIAsolution} except that we have not accounted for the type IIA dual of the gravitational backreaction of fluxes, that is, warping.
    
	Furthermore, we get
	\begin{equation}\label{eq:dJ_H_IIA}
		dJ=t\langle \vec{p},\vec{k} \rangle  \beta^0\, ,\quad H_3=b \langle \vec{p},\vec{k} \rangle \beta^0\, ,
	\end{equation}
	and the NSNS superpotential induced by $\widetilde{H}_3$ in the type IIB picture is equivalent to\footnote{One uses the relation $e^{-\phi_A}\frac{\sqrt{8\text{Vol}(X)}}{||\Omega||}=e^{-\phi_B}$.}
	\begin{equation}
		W\supset -\tau z^ak_a =\int_X dJ_{\mathbb{C}}^{(k)}\wedge \Omega_c\, ,\quad \Omega_c:=e^{-\phi_A}\Omega+iC_3\, .
	\end{equation}
	As both $J_{\mathbb{C}}$ and $\Omega$ are globally defined on $X_k$, the forms $\beta^0$ and $K$ are exact, leading to  the isomorphisms
	\begin{equation}
		H^4(X_k)\simeq H^4(X)/K\, ,\quad H^2(X_k)\simeq \left\{w\in H^2(X):\, \langle w,K\rangle =0  \right\}\, ,
	\end{equation}
	and
	\begin{equation}
		H^3(X_k)\simeq \left\{ w\in H^3(X):\,  \langle w,[\Sigma_3]^0\rangle=0\right\}/[\Sigma_3]^0\, .
	\end{equation}
	Thus, we have $b^2(X_k)=h^{1,1}(X)-1$ and $b^3(X_k)=b^3(X)-2$. In particular, as in the toroidal models, the SYZ fibration has no ``zero section'', and the generic fiber is a torsion element in homology.
	
	The relations \eqref{eq:dOmega_IIA} and \eqref{eq:dJ_H_IIA} imply that $X_k$ is \emph{half-flat} \cite{Gurrieri:2002wz}: in other words, $X_k$ is neither K\"ahler nor complex, but there still exist a globally defined and nowhere vanishing $(1,1)$-form $J$ and a $(3,0)$-form $\Omega$, neither of which is closed, but which satisfy the weaker conditions
	\begin{equation}
		d\left(\frac{1}{2}J\wedge J\right)=0\, ,\quad d\left(\text{Im}(\Omega)\right)=0\, .
	\end{equation}
	Further restricting to the PFV configurations of \cite{Demirtas:2019sip}, we get $\langle \vec{p},\vec{k}\rangle = 0$, and therefore
	\begin{equation}
		dJ = 0\, ,\quad H_3= 0\, ,
	\end{equation}
    and in particular $W_1=0$.
    
	Thus, along the vacuum locus \eqref{eq:PFV_locus} the type IIB Calabi-Yau threefold $\widetilde{X}$ equipped with $\widetilde{H}_3$ flux has a mirror-dual interpretation in type IIA string theory as an \emph{almost K\"ahler} manifold $X_k$ of $SU(3)$ structure, and with vanishing NSNS field strength.

    The non-trivial twist of the SYZ fibration of $X_k$ of course also affects how the type IIB RR-flux $F_3$ dualizes into the type IIA RR-fluxes. Instead of \eqref{eq:PFV_IIA_fluxes}, one obtains
	\begin{equation}
		F_2\supset F_2^{(A)}= -m^a \omega_a\, .
	\end{equation} 
	This modification preserves the primitivity condition \eqref{eq:F2_primitivity}, but absent orientifold planes it implies that the Bianchi identity of $F_2$ is violated, i.e., 
	\begin{equation}
		dF_2^{(A)}=-\langle \vec{m},\vec{k} \rangle  \beta^0=2Q \beta^0\, ,
	\end{equation}
	whenever the flux induced D3 charge in the type IIB frame $Q$ is non-zero. Thus, an obstruction to satisfy the $F_5$ Bianchi identity \eqref{eq:RR5_Bianchi} maps to an obstruction to the $F_2$ Bianchi identity, as it should.

    In order to arrive at the full mirror dual type IIA description one needs to incorporate orientifold planes, D-branes, and include gravitational backreaction in the form of warping, as well as the running of the type IIA dilaton. Making these backgrounds fully explicit beyond the toroidal models we discussed in \S \ref{sec:toroidalPFVs} is a formidable task, but we see no obstruction for this to work:
    
    The mirror dual of an O3/O7 type orientifold is an anti-holomorphic O6 orientifold in type IIA string theory \cite{Brunner:2003zm}, and both mobile D3-branes as well as D7-branes map to D6-branes on the type IIA side. Incorporating also the non-trivial warp factor and dilaton profiles we still expect to end up with an $SU(3)$ structure background that satisfies
    \begin{align}\label{eq:typeIIA_torsion_general}
        d\Omega&=\overline{W}_5\wedge \Omega+W_2\wedge J\, ,\quad dJ=0\, ,\\
        W_2&=-e^{\phi_A}F_2^{(8)}\, ,\quad \overline{W}_5=\bar{\del}A\, ,
    \end{align}
    and an RR two-form
    \begin{align}\label{eq:typeIIA_RRflux_general}
        F_2=&F_2|_{(1,1)}+(F_2|_{(2,0)}+c.c.)\, ,\\
        F_2|_{(1,1)}=&F_2^{(8)}\, ,\quad \left(F_2|_{(2,0)}\right)_{\gamma\delta}=g^{\alpha\bar{\beta}} \left(F_2^{(\overline{3})}\right)_{\bar{\beta}}\Omega_{\alpha \gamma \delta }^{(k)}\, ,\quad F_2^{(\overline{3})}=2i e^{-\phi_A}\bar{\del}A \, ,
    \end{align}
    with $F_2^{(8)}$ a primitive $(1,1)$ form, and dilaton profile related to the warp factor by $\phi_A-3A=const$. Therefore, gravitational backreaction is expected to deform $X_k$ from an almost-K\"ahler threefold to only a symplectic threefold. This is the mirror-dual of the backreaction induced deformation of $\tilde{X}$ from Calabi-Yau to conformal Calabi-Yau \cite{Giddings:2001yu} on the type IIB side.
    
    We emphasize that constructing the dual type IIA and M-theory backgrounds \emph{explicitly} will be a difficult task that is beyond the scope of the present work. Fortunately, in the LCS and large volume limit we can be slightly more explicit, by following essentially the same steps as in \S \ref{sec:toroidalPFVs}. After smearing all sources of D3-brane charge along the SYZ fiber, one can T-dualize the solution \eqref{eq:CY_metric_LCS} to obtain the 10d metric
    \begin{equation}\label{eq:metric_LCS}
        ds^2=e^{2A}dx_\mu dx^\mu+e^{-2A} ds^2_{\mathcal{B}}+e^{2A}h_{\alpha\beta} w^\alpha \otimes w^\beta\, ,
    \end{equation}
    where $e^{-4A}\equiv \mathcal{V}_p^{\frac{1}{3}}/\tilde{\mathcal{V}}^{\frac{2}{3}} \chi$, and $\chi$ is the solution to the electro-static problem on the (unit-volume) SYZ base. Dualizing the five-form field strength, we again get another contribution to $F_2$,
    \begin{equation}\label{eq:F2_LCS}
        F_2\supset F_2^{(B)}:=\hat{*}_{\mathcal{B}} d\chi\, ,
    \end{equation}
    and the full RR two-form flux therefore satisfies the equations of motion and Biachi identity.

    Away from the LCS limit, there are corrections to both \eqref{eq:metric_LCS} and \eqref{eq:F2_LCS} for two reasons:
    \begin{enumerate}
        \item The Ricci-flat metric on $X$ receives corrections to the LCS approximation \eqref{eq:CY_metric_LCS}.
        \item In general, the mirror-dual O6-planes do not all simply wrap the SYZ-fiber. Thus, away from the LCS limit the metric and RR two-form flux can longer be given simply in terms of a solution to a 3d electro-static problem on the base.
    \end{enumerate}
    Nevertheless, we expect the much more general relations \eqref{eq:typeIIA_torsion_general} and \eqref{eq:typeIIA_RRflux_general} of \cite{Grana:2004bg} to persist. 

    \subsection{Summary of type IIA vacua and M-theory uplift}\label{subsec:GeneralCYcase:MtheoryUplift}
    We summarize the key properties of the mirror dual type IIA background, and how the PFV conditions of \S\ref{subsec:PFVs:PFVs} get reinterpreted on the type IIA side:
    \begin{itemize}
        \item The mirror dual geometric background $X_k$ is a symplectic $SU(3)$-structure manifold. The PFV condition $\langle \vec{k},\vec{p}\rangle=0$ ensures that the torsion classes $W_1$, $W_3$ and $W_4$ vanish.
        \item The NSNS field strength on $X_k$ vanishes, again due to the PFV condition $\langle \vec{k},\vec{p}\rangle=0$.
        \item The mirror dual RR-background consists (for vanishing B-field) simply of RR two-form flux $F_2$. The integrality conditions \eqref{eq:integrality_conditions} ensure that this is consistent with Dirac quantization. The PFV condition $N_{ab}p^a p^b=k_a p^b=0$ ensures that $F_2$ is primitive.
        \item The relation $k_a=N_{ab}p^b=\kappa_{ab}m^b$ relates the $(1,1)$-component of $F_2$ to the second torsion class $W_2$, matching precisely the conditions for unbroken supersymmetry derived in \cite{Grana:2004bg}. 
    \end{itemize}
    These are the general conditions for admitting an M-theory lift on a $G_2$-manifold, and in the LCS approximation we can again write down, though somewhat implicitly, the $G_2$-metric,
    \begin{equation}\label{eq:G2_metric_generalCY_LCS}
        ds^2_M= \alpha^{-\frac{2}{3}} ds^2_{\mathbb{R}^{1,3}}
        +t^{\frac{2}{3}} \left[\mathcal{V}_p^{\frac{1}{3}}\left(\frac{(d\psi-C_1)^2}{\chi(y;y_{D6})}+ \chi(y;y_{D6}) d\hat{s}^2_{\mathcal{B}}\right)+\hat{h}_{\alpha\beta}w^\alpha\otimes w^\beta\right]\, ,
    \end{equation}
    with $d\hat{s}^2_{\mathcal{B}}$ the unit volume metric on the SYZ base of $X$, and $\hat{h}_{\alpha\beta}$ the unit-volume metric on the fiber, and with $\chi$ a solution to the suitable electro-static problem in the base (with metric $d\hat{s}^2_{\mathcal{B}}$). 
    
    Starting from a PFV of a Calabi-Yau orientifold of O3/O7 type, with orientifold-graded Hodge numbers $\tilde{h}^{1,1}_{\pm}$ and $\tilde{h}^{2,1}_{\pm}$ and $N$ mobile D3-branes, the resulting $G_2$-manifold should have Betti numbers
    \begin{equation}\label{eq:Betti_numbers_G2_general}
        b_3=b_4= \tilde{h}^{1,1}+1+3N\, ,\quad b_2=b_5=\tilde{h}^{2,1}_++N\, .
    \end{equation}
    To see this, we recall from \cite{Grimm:2004ua} that in the closed string sector of O3/O7 orientifolds one has $\tilde{h}^{1,1}_+$ K\"ahler moduli, $\tilde{h}^{1,1}_-$ torus valued two-form axions, $\tilde{h}^{2,1}_-$ complex structure moduli, and the axio-dilaton, all of which are chiral multiplets, plus $\tilde{h}^{2,1}_+$ vector multiplets. Each mobile D3-brane contributes another vector multiplet as well as three chiral multiplets containing the position moduli. The (quasi-)moduli space of PFVs is thus parameterized by $\tilde{h}^{1,1}_++\tilde{h}^{1,1}_-+1+3N=\tilde{h}^{1,1}+1+3N$ chiral multiplets, and there are $\tilde{h}^{2,1}_++N$ abelian vector multiplets, justifying  \eqref{eq:Betti_numbers_G2_general}.
    
    A caveat is the possible presence of additional moduli from seven-branes: in arriving at \eqref{eq:Betti_numbers_G2_general} we have assumed that all seven-brane gauge sectors are confining $O(8)$ Yang-Mills groups, leading to no additional massless degrees of freedom. If some D7-moduli on the type IIB side were left unstabilized, one would get additional contributions to \eqref{eq:Betti_numbers_G2_general}.

    Any confining $O(8)$ sector on seven-branes on the type IIB side should get mapped to a non-deformable singularity of the $G_2$-manifold at codimension six \cite{Acharya:2001gy}. Indeed, non-deformable such singularities giving gauge group $O(8)$ appear in K3-fibered $G_2$-manifolds constructed in \cite{Braun:2017uku}, where the K3-fiber develops a $D_4$-singularity at codimension two in the base. 
    Here, we expect the seven-brane sectors to get mapped to stacks of four D6-branes on top of an O6-plane on the type IIA side, wrapping some rigid three-cycle $\Sigma_3$. In the M-theory uplift, a local neighborhood of $\Sigma_3$ in $Y_7$ should be isomorphic to an ALE-fibration over $\Sigma_3$ that degenerates to $\mathbb{C}^2/\Gamma_{D_4}$ over codimension two loci on $\Sigma_3$ (see \cite{Pantev:2009de}). It would be interesting to explore the appearance of non-Higgsable gauge groups in these models explicitly, both from the perspective of type IIA string and in M-theory. Moreover, while in the setup of \cite{Demirtas:2019sip,Demirtas:2021nlu}, the worldvolume fluxes on seven-branes vanish and thus no chiral matter arises, it would be interesting to explore generalizations of PFVs that allow for chiral matter. If these exist they would plausibly be dual to M-theory on $G_2$-manifolds with isolated singularities (as in the non-compact examples of \cite{Acharya:2001gy}).

    Finally, let us briefly discuss the origin of the exponentially suppressed superpotential terms from the M-theory perspective. First, from the type IIB perspective, there are exponentially small \emph{classical} corrections to the otherwise quadratic superpotential (see \eqref{eq:PFV_LCS_corrections}). From the type IIA perspective, these are interpreted as euclidean worldsheets wrapped on two-cycles $\Sigma_2$ in $X_k$, which contribute as
    \begin{equation}
        W\supset -\frac{1}{(2\pi)^2}\sum_{[\Sigma_2]}\hat{n}^0_{\Sigma_2}\times \left(\int_{\Sigma}F_2\right)\times  \, e^{-2\pi \text{Vol}(\Sigma_2) +2\pi i \int_{\Sigma_2}B_2}\, ,
    \end{equation}
    where $\hat{n}_{\Sigma_2}^0$ is the genus zero Gromov-Witten invariant of the curve class $[\Sigma]$ in the underlying Calabi-Yau threefold $X$. From the M-theory perspective in turn, these are interpreted as euclidean M2-branes wrapping rigid associative three-cycles $\Sigma_3$, and the normalization of the non-perturbative superpotential contribution is equal to the \emph{Ray-Singer torsion} $T(\Sigma_3)$ \cite{Harvey:1999as}. Thus, we have the identification
    \begin{equation}\label{eq:RaySinger_toIIA}
        T(\Sigma_3)\equiv \hat{n}^0_{\Sigma_2} \int_{\Sigma_2}F_2\, .
    \end{equation}

    The other source of exponential corrections to $W$ (from the type IIB perspective) are D3 instantons and gaugino condensation contributions from confining seven-brane gauge sectors. These are mapped to D2 instanton effects (and gaugino condensation on six-branes) in type IIA string theory. In M-theory these superpotential contributions are hence also interpreted as M2-brane instanton effects.\footnote{In M-theory we expect to be able to interpret gaugino condensation contributions to the superpotential simply as M2-brane instanton effects, where the wrapped associative three-cycle is roughly the base of a local ALE fibration. I thank Jim Halverson for a useful discussion about this point.} While the precise computation of D-instanton effects in type II string theories is notoriously challenging (see \cite{Kim:2022jvv,Alexandrov:2022mmy,Kim:2022uni,Jefferson:2022ssj,Kim:2023cbh} for recent progress), here one can instead compute Ray-Singer torsions in M-theory.
    
    \section{Regimes of parametric control}
    \label{sec:ParametricControl}
    
    We have so far argued that type IIB PFVs admit geometric dual descriptions in both type IIA string theory as well as M-theory. Provided the exponential corrections to the superpotential can be taken to be parametrically small, we expect to be able to freely interpolate between the three dual descriptions by varying, say, the type IIB string coupling $e^{-\phi_B}\equiv t$ and the type IIB volume modulus. In the following, we will argue that this is indeed the case. 
    
    We will consider the following family of parametric limits:
    \begin{equation}\label{eq:limits}
        e^{-\phi_B}\rightarrow \lambda e^{-\phi_B}\, ,   \quad   \mathcal{V}\rightarrow \lambda^3 \mathcal{V}\, ,\quad \tilde{\mathcal{V}}\rightarrow \lambda^{3k} \tilde{\mathcal{V}}\, ,\quad \lambda\rightarrow \infty\, ,
    \end{equation}
    with $e^{\phi_B}$ the type IIB string coupling, $\tilde{\mathcal{V}}$ the type IIB Calabi-Yau volume (measured in string-frame) and with $\mathcal{V}$ the type IIA compactification volume (again measured in string-frame). Here, $k$ is some real parameter that parameterizes our family of limits.

    The exponential corrections to the superpotential come from classical but subleading contributions to the GVW superpotential, and scale like
    \begin{equation}
        \delta W^{RR}=\mathcal{O}( e^{-S_{LCS}})\, ,\quad S_{LCS}\sim \lambda\rightarrow \infty\, ,
    \end{equation}
    as well as D(-1) and D3 instanton corrections\footnote{Note however that D(-1) instanton corrections have been argued to be strictly absent when all D7-branes are on top of O7-planes \cite{Kim:2022jvv}. But whether or not they contribute will be irrelevant to the present discussion.} (and strong gauge dynamics), which scale as
    \begin{equation}
        \delta W^{D}=\mathcal{O}( e^{-S_{D(-1)}}, e^{-S_{D(3)}})\, ,\quad S_{D(-1)}\sim \lambda\, ,\quad S_{D3}\sim \lambda^{1+2k}\, .
    \end{equation}
    Thus, as long as $k>-\frac{1}{2}$, the superpotential decays exponentially.
    
    \subsection{IIB Regime}\label{subsec:ParametricControlIIB}
    In the limits \eqref{eq:limits}, the type IIB string coupling is parametrically small. Thus, the type IIB duality frame is parametrically controlled if all string-frame cycle volumes are parametrically large. The even-dimensional cycle volumes scale like powers of $\tilde{\mathcal{V}}$ and so we need
    \begin{equation}
        k>0\, .
    \end{equation}
    However, a stronger constraint arises from considering three-cycle volumes. These scale like
    \begin{equation}\label{eq:odd_cycle_volumesIIB}
        \text{Vol}_{IIB}(\Sigma_3^{(p)})\sim \sqrt{\tilde{\mathcal{V}}}\mathcal{V}^{\frac{p}{3}-\frac{1}{2}}\sim \lambda^{\frac{3}{2}(k-1)+p}\, ,\quad p\in \{0,1,2,3\}\, ,
    \end{equation}
    with $p$ depending on how many SYZ-fibers are wrapped (e.g., $p=0$ is the SYZ fiber and $p=3$ is the base). Thus, in order for these volumes to grow large, we actually need that
    \begin{equation}
        k>1\, .
    \end{equation}
    In the boundary case $k=1$ the SYZ fiber volume remains parametrically $\mathcal{O}(1)$.
	
	\subsection{IIA Regime}\label{subsec:ParametricControlIIA}
    For $k<1$ we see from \eqref{eq:odd_cycle_volumesIIB} that the type IIB SYZ fiber becomes parametrically small, and thus the type IIA description is expected to become controlled. Indeed, the even dimensional string-frame cycle volumes scale like positive powers of $\mathcal{V}$, and thus automatically grow large in the limit \eqref{eq:limits}, while the odd-dimensional cycle volumes now scale like
    \begin{equation}\label{eq:odd_cycle_volumesIIA}
        \text{Vol}_{IIA}(\Sigma_3^{(p)})\sim \sqrt{\mathcal{V}}\tilde{\mathcal{V}}^{\frac{p}{3}-\frac{1}{2}}\sim \lambda^{\frac{3}{2}(1-k)+pk}\, ,\quad p\in \{0,1,2,3\}\, ,
    \end{equation}
    and the type IIA string coupling scales as
    \begin{equation}
        e^{-\phi_A}=e^{-\phi_B}\text{Vol}_{IIB}(\Sigma_3^{(0)})\sim \lambda^{\frac{1}{2}(3k-1)}\, .
    \end{equation}
    Thus, the type IIA background is weakly coupled, and all string-frame volumes are parametrically large if
    \begin{equation}
        \frac{1}{3}<k<1\, .
    \end{equation}
    In the boundary case $k=\frac{1}{3}$ the type IIA string coupling remains parametrically of $\mathcal{O}(1)$.
	
	\subsection{M-theory Regime}\label{subsec:ParametricControlMtheory}
    For $k<\frac{1}{3}$ we thus conclude that the M-theory description becomes the appropriate one. Indeed, the M-theory circle radius, as measured in 11d Planck units, scales as
    \begin{equation}
        \text{Vol}_M(S^1_M)=e^{\frac{2}{3}\phi_A}\sim \lambda^{\frac{1}{3}-k}\, ,
    \end{equation}
    and thus grows large when $k<1/3$.
    
    Moreover, the even-dimensional cycles in the base of the circle fibration scale as positive powers of
    \begin{equation}
        \text{Vol}_M(\Sigma_2)= e^{-\frac{2}{3}\phi_A}\mathcal{V}^{\frac{1}{3}}\sim  \lambda^{k+\frac{2}{3}}\, ,
    \end{equation}
    while the odd-dimensional cycles in the base scale like
    \begin{equation}
        \text{Vol}_{M}(\Sigma_3^{(p)})=e^{-\phi_A}\text{Vol}_{IIA}(\Sigma_3^{(p)})\sim \lambda^{1+pk}\, , \quad p\in \{0,1,2,3\}\, .
    \end{equation}
    All of these grow parametrically large if $k>-\frac{1}{3}$.

    There remains the window of limits 
    \begin{equation}
        -\frac{1}{2}<k\leq -\frac{1}{3}\, ,
    \end{equation}
    in which the superpotential still decays exponentially, but  the volume of the SYZ-base appears to vanish: $\text{Vol}_{M}(\Sigma_3^{(3)})\sim \lambda^{1+3k}\rightarrow 0$. 
    
    However, we think the M-theory description actually remains weakly curved as long as $k>-1/2$.
    First, we recall from the discussion in \S\ref{sec:toroidalPFVs} and \S\ref{sec:GeneralCYcase} that the twist of the SYZ fibration from T-dualizing the type IIB NSNS fluxes removes the SYZ base as a sub-manifold, and so there actually exists no calibrated cycle whose volume scales like $\lambda^{1+3k}$. Nevertheless, as long as the M-theory compactification can be thought of as a $T^4$-fibration over the base of the SYZ fibration of $X$ one would expect curvature invariants to blow up when the base volume shrinks. But, in a limit \eqref{eq:limits} with $k<0$ the SYZ-fiber of $X$ grows faster than the base. In this regime, we no longer expect $X$ to be $T^3$-fibered, and similarly we do not expect $Y_7$ to remain $T^4$-fibered. 
    
    Indeed, in the explicit solutions of \S \ref{sec:toroidalPFVs}, which na\"ively are manifestly $T^4$-fibered, the constant term $c$ in the solution for the warp factor \eqref{eq:chi_solution} scales like a type IIB string-frame four-cycle volume,
    \begin{equation}
        c\sim \lambda^{2k}\, ,
    \end{equation}
    and so vanishes in the limits \eqref{eq:limits} for $k<0$. Therefore, in these limits the metric \eqref{eq:G2_metric} (and its generalization \eqref{eq:G2_metric_generalCY_LCS}) seizes to well-approximate the $G_2$-metric, and we see no clear evidence for a diverging curvature invariant.
    
\begin{figure}
        \centering
        \includegraphics[width=0.6\linewidth]{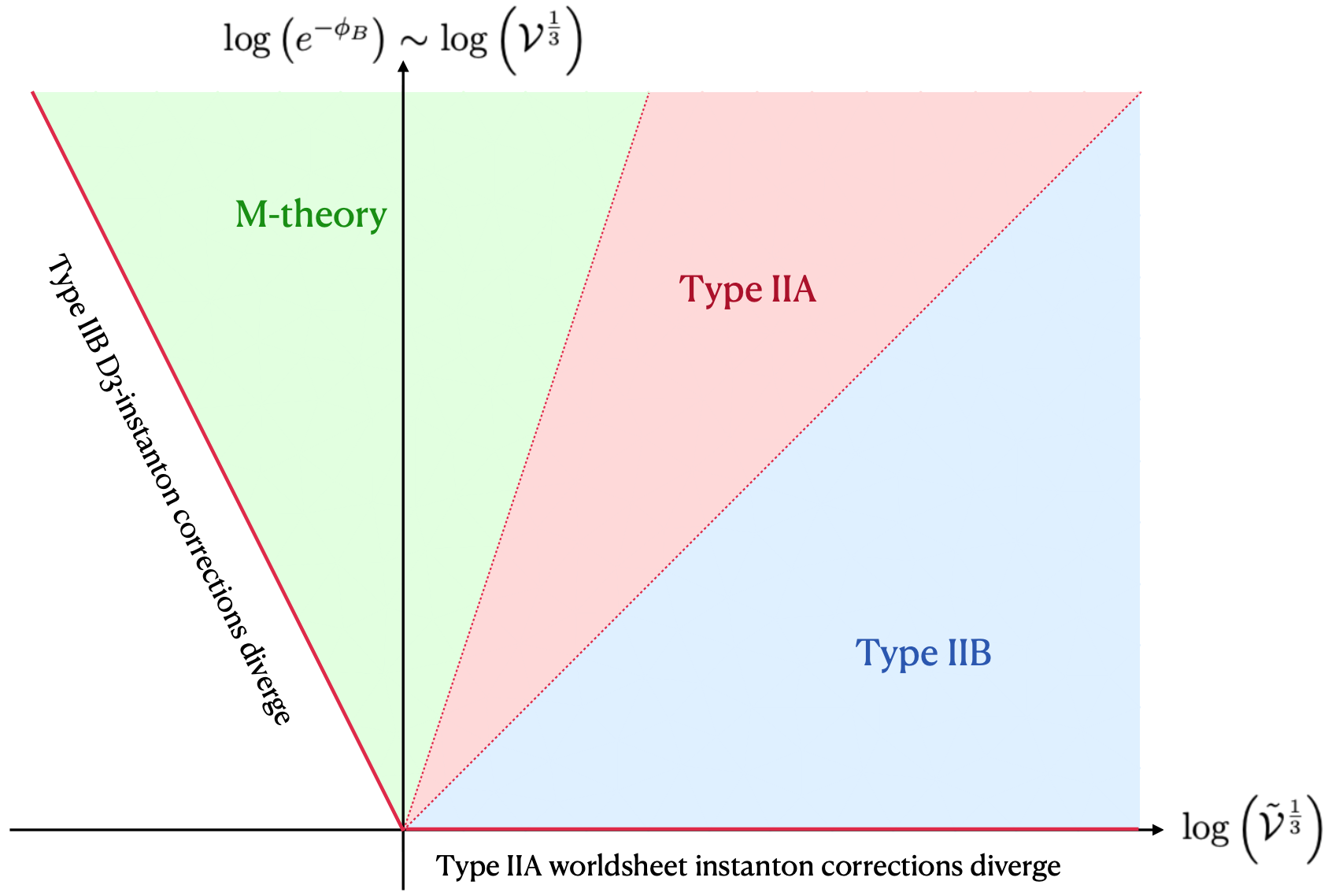}
        \caption{The regimes of parametric control in type IIB string theory (blue), type IIA string theory (red), and M-theory (green). To the upper right of the solid red lines, the superpotential is exponentially small. Near the solid red lines the exponential corrections to the superpotential become $\mathcal{O}(1)$. The dashed lines mark the transitions between controlled Type IIB and Type IIA description (with $\mathcal{O}(1)$ SYZ-fiber volume), and between Type IIA and M-theory (with $\mathcal{O}(1)$ string coupling).}
        \label{fig:smallW_triality}
    \end{figure}

    We find it natural to conjecture that the M-theory description remains weakly curved provided the smallest calibrated cycle volume is large. In our setup, this volume scales like $\lambda^{1+2k}$, and so we expect the M-theory description to remain valid in the full window
    \begin{equation}
        -\frac{1}{2}< k<\frac{1}{3}\, .
    \end{equation}
    Combining the three duality frames we are then able to cover the entire range of limits \eqref{eq:limits} in which the superpotential decays exponentially! See Figure \ref{fig:smallW_triality} for an illustration of the three regions of parametric control.

    Finally, recalling that the vacua of \cite{Demirtas:2019sip,Demirtas:2021nlu} live in the regime
    \begin{equation}
        \text{Re}(T_i)\sim e^{-\phi_B}\tilde{\mathcal{V}}^{\frac{2}{3}}\sim e^{-\phi_B}\sim \lambda:=\frac{\log(1/|W_0|)}{2\pi}\gg 1\, ,
    \end{equation}
    we see that the limit $W_0\rightarrow 0$ can be thought of as a member of our family of limits \eqref{eq:limits} with $k=0$.
    This particular limit is quite special because all length scales $L$ grow homogeneously as 
    \begin{equation}
        L\sim \lambda^{\frac{1}{3}}\rightarrow \infty\, .
    \end{equation}
    As a consequence we are led to conclude that the limit of exponentially small flux superpotential in the type IIB description in the vacua of \cite{Demirtas:2019sip,Demirtas:2021nlu} is more naturally interpreted as the universal large volume limit of a $G_2$-compactification in M-theory! 
    
    In hindsight this result is easy to understand: the mechanism of \cite{Demirtas:2019sip} is of racetrack type, and thus involves balancing exponential terms in the superpotential against each other. As all the exponential corrections involved in this setup are interpreted as M2-instantons in M-theory (see our discussion at the end of \S\ref{subsec:GeneralCYcase:MtheoryUplift}), it follows that a basis of smallest calibrated three-cycles must have comparable volumes. If instanton actions are large we then end up with roughly isotropic large volume compactifications.

    \section{A strengthened singular bulk problem}
    \label{sec:SingularBulk}

    In this section we would like to continue discussing the status of computational control in flux compactifications of KKLT type \cite{Kachru:2003aw}, specifically in its incarnations as perturbatively flat vacua \cite{Demirtas:2019sip,Demirtas:2021nlu,McAllister:2024lnt}. Specifically, we will be concerned with a potential control problem discussed in \cite{Carta:2019rhx,Gao:2020xqh,Carta:2021lqg}, termed \emph{singular bulk problem} in \cite{Gao:2020xqh}.

    First, let us recall the issue. For this we return to the 6d respectively electrostatic problem that appeared in the construction of the ten dimensional type IIB solution, which we reproduce here for the reader's convenience:
    \begin{equation}\label{eq:6delectrostatic_copy}
		e^{-\phi_B}\nabla^2_{\widetilde{X}}e^{-4A} =\frac{2Q}{\tilde{\mathcal{V}}}+\rho^{\text{loc}}_{\text{D3}}\, ,
	\end{equation}
    where $\nabla^2_{\widetilde{X}}$ is the Laplacian constructed from the string-frame Calabi-Yau metric (i.e., without including the conformal inverse warp factor). The r.h. side is simply the D3-charge density on the Calabi-threefold. We can rephrase this as
    \begin{equation}\label{eq:IIB_rescaled_Poisson}
        \hat{\nabla}^2_{\widetilde{X}}\tilde{\chi}=2Q+\hat{\rho}_{D3}^{\text{loc}}\, ,\quad \tilde{\chi}:=e^{-\phi_B}\tilde{\mathcal{V}}^{\frac{2}{3}} e^{-4A}\, ,
    \end{equation}
    where now the charge density and Laplacian are evaluated with respect to the unit-volume Calabi-Yau metric $\hat{g}$. As in \S\ref{sec:toroidalPFVs} we can write the general solution to this Poisson problem as
    \begin{equation}\label{eq:IIB_warp_factor_sol}
        \tilde{\chi}=\tilde{\chi}_0+\tilde{c}\, ,\quad \int_{\widetilde{X}}d^6y\sqrt{\hat{g}}\,\chi_0=0\, .
    \end{equation}
    Thus by definition --- if the Calabi-Yau $\widetilde{X}$ is roughly isotropic --- for $\tilde{c}=0$ the warp factor is negative over an $\mathcal{O}(1)$ fraction of the Calabi-Yau threefold $\widetilde{X}$. 
    
    How large $\tilde{c}$ has to be before regions of negative warp factor shrink appreciably depends on the D3-charge distribution on the r.h. side of \eqref{eq:IIB_rescaled_Poisson}. Fore example, if the sources of negative D3 charge are well-separated from all positive D3-charge a strong constraint $\tilde{c}\gtrsim \mathcal{Q}$ arises (see \cite{Carta:2019rhx,Gao:2020xqh} for details). If on the other hand there are $N\gg 1$ localized sources which are randomly distributed over $\widetilde{X}$, D3 charge is partially screened over long distances and $\tilde{c}\gtrsim \mathcal{Q}/\sqrt{N}$ should suffice.
    
    In general, one should expand the D3-charge distribution into eigenfunctions of the Laplacian on $\widetilde{X}$, and 
    impose
    \begin{equation}\label{eq:SBPcontrol0}
        \tilde{c}\gtrsim Q_2\, ,
    \end{equation}
    with $Q_2$ the coefficient of the eigenfunction of smallest non-zero eigenvalue.

    Now, Einstein-frame volumes of holomorphic four-cycles $D$ are the real parts of the holomorphic K\"ahler coordinates $T_D$ \cite{Shiu:2008ry,Frey:2008xw,Martucci:2016pzt},
    \begin{equation}
        \text{Re}(T_D)=\frac{1}{2}\int_D e^{-\phi_B}e^{-4A}\tilde{J}\wedge \tilde{J} = \frac{1}{2}\int_D \tilde{\chi} \hat{\tilde{J}}\wedge \hat{\tilde{J}}\, ,
    \end{equation}
    where $\hat{\tilde{J}}$ is the K\"ahler form on $\widetilde{X}$ normalized to unit-volume $\frac{1}{3!}\int_{\widetilde{X}}\hat{\tilde{J}}^3=1$, and therefore the constant part $\tilde{c}$ in the warp factor solution \eqref{eq:IIB_warp_factor_sol} scales like an Einstein-frame divisor volume. For this reason, in \cite{Demirtas:2021nlu,McAllister:2024lnt}, 
    the overall amount of negative D3-charge on seven-branes and O3-planes ---  i.e., $\mathcal{Q}$ in \eqref{eq:D3_tadpole} --- was compared to the Einstein-frame volume $e^{-\phi_B}\tilde{\mathcal{V}}^{\frac{2}{3}}$ as a conservative estimate of the importance of warping corrections to the K\"ahler potential and K\"ahler coordinates, finding
    \begin{equation}\label{eq:SBP_bound0_conservative}
        \epsilon_{SBP}:=\frac{\mathcal{Q}}{e^{-\phi_B}\tilde{\mathcal{V}}^{\frac{2}{3}}}\ll 1\, ,
    \end{equation}
    and concluding from this (and other related control measures) that warping corrections are under decent control. This estimate is conservative because it takes into account no screening of D3-charge, i.e., $Q_2\sim \mathcal{Q}$.

    So far we have assumed that the Calabi-Yau is roughly isotropic, but all known constructions with very small flux superpotential, and hence very small cosmological constants, actually have large complex structure. In other words the SYZ-fiber is much smaller than the base. In this case, we are able to derive a new and stronger control constraint than \eqref{eq:SBPcontrol0}. For this, we will only assume that the SYZ base is roughly isotropic (in the non-isotropic case a stronger bound is expected). As the fiber is much smaller than the base, we can dimensionally reduce to seven dimensions, and inquire about the profile of the Kaluza-Klein zero mode $\tilde{\chi}_0$ of $\tilde{\chi}$ along the fiber. This zero mode acquires a profile along the base directions determined by a 3d Poisson equation, which we can obtain by averaging \eqref{eq:IIB_rescaled_Poisson} over the fiber directions,
    \begin{equation}
        \nabla^2_{\mathcal{B}} \tilde{\chi}_0=2Q+\langle \hat{\rho}_{D3}^{ \text{loc} } \rangle\, ,
    \end{equation}
    where $\langle \cdot \rangle$ denotes averaging over the fiber, and $\nabla^2_{\mathcal{B}}$ is the Laplacian on the base, for a unit volume Calabi-Yau $\widetilde{X}$. 
    
    Echoing the above discussion for the isotropic Calabi-Yau case, we should now scale out the base volume, which at large complex structure is given by $\sqrt{\mathcal{V}}$ (the fiber volume is equal to $1/\sqrt{\mathcal{V}}$). In other words, we can reformulate the Poisson problem as 
    \begin{equation}\label{eq:dangerous_Poisson}
        \hat{\nabla}^2_{\mathcal{B}} \chi=2Q+\langle \hat{\rho}_{D3}^{ \text{loc} } \rangle\, ,\quad \chi:=\frac{\tilde{\chi}_0}{\mathcal{V}^{\frac{1}{3}}}\, ,
    \end{equation}
    where $\hat{\nabla}^2_{\mathcal{B}}$ is the Laplacian on the SYZ base constructed from the unit-volume metric $\hat{g}_{\mathcal{B}}$. Again, we can write the general solution as
    \begin{equation}
        \chi=\chi_0+c\, ,\quad \int_{\mathcal{B}}d^3y\sqrt{\hat{g}_{\mathcal{B}}}\,\chi_0=0\, ,
    \end{equation}
    and the free parameter $c$ is related to the parameter $\tilde{c}$ in \eqref{eq:IIB_warp_factor_sol} via $c=\tilde{c}/\mathcal{V}^{\frac{1}{3}}$.\footnote{The function $\chi$ and parameter $c$ here are the same as the quantities with the same name in \S\ref{sec:toroidalPFVs} and \S\ref{sec:GeneralCYcase}.} 
    
    By complete analogy to the previous discussion, avoiding a large region with negative warp factor actually requires that the parameter $c$ is large compared to the typical amount of unscreened D3-charge over long distances in $\mathcal{B}$. Averaging over the SYZ fiber \emph{might} significantly reduce this quantity compared to $Q_2$. But if it does not, we should impose
    \begin{equation}\label{eq:SBPcontrol}
        c\overset{!}{\gg} Q_0\quad \longleftrightarrow\quad  \tilde{c}\overset{!}{\gg} \mathcal{V}^{\frac{1}{3}} Q_2\gg Q_2\, ,
    \end{equation}
    and at large complex structure, $\mathcal{V}\gg 1$, this is clearly a stronger constraint than \eqref{eq:SBPcontrol0}.\footnote{See \cite{Junghans:2023lpo} for a related discussion (compatible with ours) of backreaction effects in a different context.}
    
    We conclude that at large complex structure the control parameter for avoiding a singular bulk problem \eqref{eq:SBPcontrol0} becomes worse by a factor $\mathcal{V}^{\frac{1}{3}}$. In the particular case of perturbatively flat vacua we have $\mathcal{V}^{\frac{1}{3}}\sim e^{-\phi_B}$, and the conservative bound \eqref{eq:SBP_bound0_conservative} simplifies to
    \begin{equation}\label{eq:SBPcontrol_conservative}
        \frac{\mathcal{Q}}{\tilde{\mathcal{V}}^{\frac{2}{3}}}\ll 1\, .
    \end{equation}
    In other words, Einstein-frame volumes get effectively replaced by string-frame volumes in assessing warping corrections. 
    
    The bound \eqref{eq:SBPcontrol_conservative} is not comfortably satisfied in the constructions \cite{Demirtas:2021nlu,McAllister:2024lnt}, raising the possibility of uncontrolled corrections to the K\"ahler potential and K\"ahler coordinates in these models. The situation might well be less severe: in general $Q_2< \mathcal{Q}$ and \eqref{eq:SBPcontrol_conservative} might be a significant overestimate of the severity of the singular bulk problem. In order to conclusively assess this, one should solve the Poisson problem explicitly using the numerical Calabi-Yau metric in explicit examples, which is beyond the scope of this work. However, given the significant recent progress in computing numerical Calabi-Yau metrics \cite{Anderson:2020hux,Douglas:2020hpv,Jejjala:2020wcc,Larfors:2022nep,Gerdes:2022nzr,Halverson:2023ndu} this seems within reach.

    There are two possible outcomes of such an analysis: 
    \begin{enumerate}
        \item A precise assessment of the singular bulk problem in the solutions \cite{Demirtas:2021nlu} using numerical metrics reveals that warping corrections are still small. In this case, the computation of the K\"ahler potential and K\"ahler coordinates of \cite{Demirtas:2021nlu} is confirmed. Explicitly constructing the M-theory duals would be extremely interesting, but likely not necessary to control the vacua with exponentially small $W_0$ of \cite{Demirtas:2021nlu}.
        \item Warping corrections are in fact unsuppressed.
    \end{enumerate}
    In case (b), one needs to compute warping corrections explicitly, and fortunately our discussion of \S\ref{sec:ParametricControl} shows how this can actually be done in practice: one needs to construct the $G_2$-manifolds dual to the PFVs of \cite{Demirtas:2021nlu}, and compute the M-theory K\"ahler potential at large volume \cite{Beasley:2002db},
    \begin{equation}
        K=-3\log\left(\int_{Y_7}\Phi\wedge *\Phi\right)+const.\, ,
    \end{equation}
    as a function of the periods of the associative three-form $\Phi$, as done e.g. for orbifold resolutions in \cite{Lukas:2003dn}.\footnote{The usefulness of the M-theory lifts to compute warping effects has also been emphasized in \cite{Andriolo:2018yrz}.} Importantly, due to holomorphy, the computation of the relevant non-perturbative corrections to the superpotential can be carried over directly from the type II side.

    What could be the effect of incorporating unsuppressed warping corrections into the analysis of \cite{Demirtas:2021nlu}? The supersymmetric AdS vacua with exponentially small superpotential might simply shift to a new location in moduli space. Indeed, scale separation in the vacua of \cite{Demirtas:2021nlu} is driven by the  exponential smallness of the superpotential, which is unaffected by the questions at hand, and no particular structures in the K\"ahler potential appear to be required. Of course, only by actually computing the corrections explicitly could one confirm (or refute) this expectation. But we emphasize that because warping corrections can be understood in M-theory as a purely classical geometric effect, they now appear to be the (presently unknown) solution of a well-posed problem, rather than a signature of a true pathology.

    We hope to return to these issues elsewhere.

    \section{Outlook}
    \label{sec:Conclusions}

    In this paper we have studied a large class of explicit flux vacua in type IIB string theory --- the \emph{perturbatively flat vacua} (PFVs) of \cite{Demirtas:2019sip} --- which arise from enumerating solutions to certain Diophantine equations in flux quanta. We have argued that these vacua admit dual descriptions as type IIA flux compactifications on symplectic manifolds, and in particular as M-theory compactifications on $G_2$-holonomy manifolds.

    A central open problem is to construct the resulting $G_2$-manifolds explicitly, and to investigate potential connections to known constructions such as Joyce’s resolutions of orbifold singularities \cite{JoyceI,JoyceII} or twisted connected sums \cite{Kovavlev:2003,Corti:2012aa,Corti:2012kd}. This would be a significant step forward, as it would enable the systematic computation of warping corrections to the effective field theory of type II flux vacua in purely classical geometric terms (\emph{cf.} \cite{Andriolo:2018yrz}), and may even be necessary to control the Kähler potential in the vacuum constructions of \cite{Demirtas:2021nlu}. A particular target is to compute Kähler potentials for these $G_2$-compactifications along the lines of \cite{Lukas:2003dn}.

    Another promising direction concerns the computation of D-instanton effects in type II compactifications. As is well known, M2-brane instanton effects in M-theory descend to worldsheet instantons and D2-brane instantons in type IIA string theory. While worldsheet instanton corrections can be computed geometrically in the type IIB mirror, D-instanton effects are much more difficult to calculate explicitly (but see \cite{Kim:2022jvv,Alexandrov:2022mmy,Kim:2022uni,Jefferson:2022ssj,Kim:2023cbh}). A given $G_2$-compactification may, however, admit \emph{multiple} distinct type IIA limits arising from different circle fibrations \cite{Kachru:2001je}. This raises the possibility that D2-instanton effects in one such limit could be mapped to worldsheet instanton corrections in another.\footnote{See \cite{Anderson:2016cdu} for a conceptually similar phenomenon involving multiple genus-one fibrations in Calabi–Yau compactifications of M/F-theory.} In this way, certain D-instanton effects --- such as D3-instanton contributions to the superpotential in type IIB compactifications --- might be computable via classical geometry in a dual model.

    A further intriguing question is whether the type IIB mechanism of spontaneous supersymmetry breaking at the tips of warped throats \cite{Kachru:2002gs,Kachru:2003aw} admits a natural reinterpretation in M-theory (\emph{cf.} \cite{Andriolo:2018yrz}). The exponentially small complex structure modulus that characterizes a warped throat in type IIB string theory corresponds to a shrunken curve in the type IIA description, which in turn maps to a geometric singularity in the M-theory $G_2$-manifold \cite{Atiyah:2000zz,Becker:2004qh,Becker:2005ef}. This suggests that supersymmetry breaking could be understood directly in M-theory by starting with a supersymmetric singularity, and then deforming it by an exponentially small amount in a way that is incompatible with $G_2$-holonomy.

    \section*{Acknowledgements} I would like to thank Joe Davighi, Andrew Frey, Naomi Gendler, Jim Halverson, Daniel Junghans, Manki Kim, Severin L\"ust, Liam McAllister, Ruben Minasian, Ruben Monten and Ethan Torres for helpful discussions, and Manki Kim, Liam McAllister and an anonymous referee for insightful comments on a draft. I am grateful to the Erwin-Schr\"odinger Institute in Vienna for its hospitality during the Programme ``The Landscape vs. the Swampland" where this work was initiated. Support for this research was provided by the Office of the Vice Chancellor for Research and Graduate Education at the University of Wisconsin-Madison with funding from  the Wisconsin Alumni  Research Foundation.

\bibliography{biblio}

\providecommand{\href}[2]{#2}\begingroup\raggedright\begin{thebibliography}{100}

\bibitem{Demirtas:2019sip}
M.~Demirtas, M.~Kim, L.~Mcallister, and J.~Moritz, {\it {Vacua with Small Flux
  Superpotential}},  {\em Phys. Rev. Lett.} {\bf 124} (2020), no.~21 211603,
  [\href{http://arxiv.org/abs/1912.10047}{{\tt arXiv:1912.10047}}].

\bibitem{McAllister:2023vgy}
L.~McAllister and F.~Quevedo, {\it {Moduli Stabilization in String Theory}},
  \href{http://arxiv.org/abs/2310.20559}{{\tt arXiv:2310.20559}}.

\bibitem{Giddings:2001yu}
S.~B. Giddings, S.~Kachru, and J.~Polchinski, {\it {Hierarchies from fluxes in
  string compactifications}},  {\em Phys. Rev. D} {\bf 66} (2002) 106006,
  [\href{http://arxiv.org/abs/hep-th/0105097}{{\tt hep-th/0105097}}].

\bibitem{DeWolfe:2005uu}
O.~DeWolfe, A.~Giryavets, S.~Kachru, and W.~Taylor, {\it {Type IIA moduli
  stabilization}},  {\em JHEP} {\bf 07} (2005) 066,
  [\href{http://arxiv.org/abs/hep-th/0505160}{{\tt hep-th/0505160}}].

\bibitem{Dasgupta:1999ss}
K.~Dasgupta, G.~Rajesh, and S.~Sethi, {\it {M theory, orientifolds and G -
  flux}},  {\em JHEP} {\bf 08} (1999) 023,
  [\href{http://arxiv.org/abs/hep-th/9908088}{{\tt hep-th/9908088}}].

\bibitem{Beasley:2002db}
C.~Beasley and E.~Witten, {\it {A Note on fluxes and superpotentials in M
  theory compactifications on manifolds of G(2) holonomy}},  {\em JHEP} {\bf
  07} (2002) 046, [\href{http://arxiv.org/abs/hep-th/0203061}{{\tt
  hep-th/0203061}}].

\bibitem{Kachru:2003aw}
S.~Kachru, R.~Kallosh, A.~D. Linde, and S.~P. Trivedi, {\it {De Sitter vacua in
  string theory}},  {\em Phys. Rev. D} {\bf 68} (2003) 046005,
  [\href{http://arxiv.org/abs/hep-th/0301240}{{\tt hep-th/0301240}}].

\bibitem{Demirtas:2020ffz}
M.~Demirtas, M.~Kim, L.~McAllister, and J.~Moritz, {\it {Conifold Vacua with
  Small Flux Superpotential}},  {\em Fortsch. Phys.} {\bf 68} (2020) 2000085,
  [\href{http://arxiv.org/abs/2009.03312}{{\tt arXiv:2009.03312}}].

\bibitem{Alvarez-Garcia:2020pxd}
R.~\'Alvarez-Garc\'\i{}a, R.~Blumenhagen, M.~Brinkmann, and L.~Schlechter, {\it
  {Small Flux Superpotentials for Type IIB Flux Vacua Close to a Conifold}},
  {\em Fortsch. Phys.} {\bf 68} (2020) 2000088,
  [\href{http://arxiv.org/abs/2009.03325}{{\tt arXiv:2009.03325}}].

\bibitem{Demirtas:2021nlu}
M.~Demirtas, M.~Kim, L.~McAllister, J.~Moritz, and A.~Rios-Tascon, {\it {Small
  cosmological constants in string theory}},  {\em JHEP} {\bf 12} (2021) 136,
  [\href{http://arxiv.org/abs/2107.09064}{{\tt arXiv:2107.09064}}].

\bibitem{McAllister:2024lnt}
L.~McAllister, J.~Moritz, R.~Nally, and A.~Schachner, {\it {Candidate de Sitter
  vacua}},  {\em Phys. Rev. D} {\bf 111} (2025), no.~8 086015,
  [\href{http://arxiv.org/abs/2406.13751}{{\tt arXiv:2406.13751}}].

\bibitem{Moritz:2017xto}
J.~Moritz, A.~Retolaza, and A.~Westphal, {\it {Toward de Sitter space from ten
  dimensions}},  {\em Phys. Rev. D} {\bf 97} (2018), no.~4 046010,
  [\href{http://arxiv.org/abs/1707.08678}{{\tt arXiv:1707.08678}}].

\bibitem{Gautason:2018gln}
F.~F. Gautason, V.~Van~Hemelryck, and T.~Van~Riet, {\it {The Tension between
  10D Supergravity and dS Uplifts}},  {\em Fortsch. Phys.} {\bf 67} (2019),
  no.~1-2 1800091, [\href{http://arxiv.org/abs/1810.08518}{{\tt
  arXiv:1810.08518}}].

\bibitem{Hamada:2018qef}
Y.~Hamada, A.~Hebecker, G.~Shiu, and P.~Soler, {\it {On brane gaugino
  condensates in 10d}},  {\em JHEP} {\bf 04} (2019) 008,
  [\href{http://arxiv.org/abs/1812.06097}{{\tt arXiv:1812.06097}}].

\bibitem{Hamada:2019ack}
Y.~Hamada, A.~Hebecker, G.~Shiu, and P.~Soler, {\it {Understanding KKLT from a
  10d perspective}},  {\em JHEP} {\bf 06} (2019) 019,
  [\href{http://arxiv.org/abs/1902.01410}{{\tt arXiv:1902.01410}}].

\bibitem{Carta:2019rhx}
F.~Carta, J.~Moritz, and A.~Westphal, {\it {Gaugino condensation and small
  uplifts in KKLT}},  {\em JHEP} {\bf 08} (2019) 141,
  [\href{http://arxiv.org/abs/1902.01412}{{\tt arXiv:1902.01412}}].

\bibitem{Gautason:2019jwq}
F.~F. Gautason, V.~Van~Hemelryck, T.~Van~Riet, and V.~Venken, {\it {A 10d view
  on the KKLT AdS vacuum and uplifting}},  {\em JHEP} {\bf 06} (2020) 074,
  [\href{http://arxiv.org/abs/1902.01415}{{\tt arXiv:1902.01415}}].

\bibitem{Bena:2019mte}
I.~Bena, M.~Gra\~na, N.~Kovensky, and A.~Retolaza, {\it {K\"ahler moduli
  stabilization from ten dimensions}},  {\em JHEP} {\bf 10} (2019) 200,
  [\href{http://arxiv.org/abs/1908.01785}{{\tt arXiv:1908.01785}}].

\bibitem{Kachru:2019dvo}
S.~Kachru, M.~Kim, L.~Mcallister, and M.~Zimet, {\it {de Sitter vacua from ten
  dimensions}},  {\em JHEP} {\bf 12} (2021) 111,
  [\href{http://arxiv.org/abs/1908.04788}{{\tt arXiv:1908.04788}}].

\bibitem{Randall:2019ent}
L.~Randall, {\it {The Boundaries of KKLT}},  {\em Fortsch. Phys.} {\bf 68}
  (2020), no.~3-4 1900105, [\href{http://arxiv.org/abs/1912.06693}{{\tt
  arXiv:1912.06693}}].

\bibitem{Bena:2020xrh}
I.~Bena, J.~Bl\r{a}b\"ack, M.~Gra\~na, and S.~L\"ust, {\it {The tadpole
  problem}},  {\em JHEP} {\bf 11} (2021) 223,
  [\href{http://arxiv.org/abs/2010.10519}{{\tt arXiv:2010.10519}}].

\bibitem{Lust:2022lfc}
S.~L\"ust, C.~Vafa, M.~Wiesner, and K.~Xu, {\it {Holography and the KKLT
  scenario}},  {\em JHEP} {\bf 10} (2022) 188,
  [\href{http://arxiv.org/abs/2204.07171}{{\tt arXiv:2204.07171}}].

\bibitem{Lust:2022xoq}
S.~L\"ust and L.~Randall, {\it {Effective Theory of Warped Compactifications
  and the Implications for KKLT}},  {\em Fortsch. Phys.} {\bf 70} (2022),
  no.~7-8 2200103, [\href{http://arxiv.org/abs/2206.04708}{{\tt
  arXiv:2206.04708}}].

\bibitem{Bena:2024are}
I.~Bena, Y.~Li, and S.~L\"ust, {\it {KKLT Ex Nihilo}},
  \href{http://arxiv.org/abs/2410.22400}{{\tt arXiv:2410.22400}}.

\bibitem{Grana:2004bg}
M.~Grana, R.~Minasian, M.~Petrini, and A.~Tomasiello, {\it {Supersymmetric
  backgrounds from generalized Calabi-Yau manifolds}},  {\em JHEP} {\bf 08}
  (2004) 046, [\href{http://arxiv.org/abs/hep-th/0406137}{{\tt
  hep-th/0406137}}].

\bibitem{VanHemelryck:2024bas}
V.~Van~Hemelryck, {\it {Weak G2 manifolds and scale separation in M-theory from
  type IIA backgrounds}},  {\em Phys. Rev. D} {\bf 110} (2024), no.~10 106013,
  [\href{http://arxiv.org/abs/2408.16609}{{\tt arXiv:2408.16609}}].

\bibitem{JoyceI}
D.~Joyce, {\it Compact riemannian {7}-manifolds with holonomy $g\sb 2$. i},
  {\em Journal of Differential Geometry - J DIFFEREN GEOM} {\bf 43} (01, 1996).

\bibitem{JoyceII}
D.~Joyce, {\it Compact riemannian 7-manifolds with holonomy $g\sb 2$. ii},
  {\em Journal of Differential Geometry} {\bf 43} (01, 1996) 329--375.

\bibitem{Kovavlev:2003}
A.~Kovalev, {\it Twisted connected sums and special riemannian holonomy},  {\em
  Journal für die reine und angewandte Mathematik} {\bf 2003} (2003), no.~565
  125--160.

\bibitem{Corti:2012aa}
A.~{Corti}, M.~{Haskins}, J.~{Nordstr{\"o}m}, and T.~{Pacini}, {\it
  {Asymptotically cylindrical Calabi-Yau 3-folds from weak Fano 3-folds}},
  {\em arXiv e-prints} (June, 2012) arXiv:1206.2277,
  [\href{http://arxiv.org/abs/1206.2277}{{\tt arXiv:1206.2277}}].

\bibitem{Corti:2012kd}
A.~Corti, M.~Haskins, J.~Nordstr\"om, and T.~Pacini, {\it
  {$\mathrm{G}_{2}$-manifolds and associative submanifolds via semi-Fano
  $3$-folds}},  {\em Duke Math. J.} {\bf 164} (2015), no.~10 1971--2092,
  [\href{http://arxiv.org/abs/1207.4470}{{\tt arXiv:1207.4470}}].

\bibitem{Halverson:2014tya}
J.~Halverson and D.~R. Morrison, {\it {The landscape of M-theory
  compactifications on seven-manifolds with G$_{2}$ holonomy}},  {\em JHEP}
  {\bf 04} (2015) 047, [\href{http://arxiv.org/abs/1412.4123}{{\tt
  arXiv:1412.4123}}].

\bibitem{Braun:2017uku}
A.~P. Braun and S.~Sch\"afer-Nameki, {\it {Compact, Singular $G_2$-Holonomy
  Manifolds and M/Heterotic/F-Theory Duality}},  {\em JHEP} {\bf 04} (2018)
  126, [\href{http://arxiv.org/abs/1708.07215}{{\tt arXiv:1708.07215}}].

\bibitem{Braun:2018fdp}
A.~P. Braun, M.~Del~Zotto, J.~Halverson, M.~Larfors, D.~R. Morrison, and
  S.~Sch\"afer-Nameki, {\it {Infinitely many M2-instanton corrections to
  M-theory on G$_{2}$-manifolds}},  {\em JHEP} {\bf 09} (2018) 077,
  [\href{http://arxiv.org/abs/1803.02343}{{\tt arXiv:1803.02343}}].

\bibitem{Kachru:2002sk}
S.~Kachru, M.~B. Schulz, P.~K. Tripathy, and S.~P. Trivedi, {\it {New
  supersymmetric string compactifications}},  {\em JHEP} {\bf 03} (2003) 061,
  [\href{http://arxiv.org/abs/hep-th/0211182}{{\tt hep-th/0211182}}].

\bibitem{Grana:2006kf}
M.~Grana, R.~Minasian, M.~Petrini, and A.~Tomasiello, {\it {A Scan for new N=1
  vacua on twisted tori}},  {\em JHEP} {\bf 05} (2007) 031,
  [\href{http://arxiv.org/abs/hep-th/0609124}{{\tt hep-th/0609124}}].

\bibitem{Andriolo:2018yrz}
S.~Andriolo, G.~Shiu, H.~Triendl, T.~Van~Riet, V.~Venken, and G.~Zoccarato,
  {\it {Compact G2 holonomy spaces from SU(3) structures}},  {\em JHEP} {\bf
  03} (2019) 059, [\href{http://arxiv.org/abs/1811.00063}{{\tt
  arXiv:1811.00063}}].

\bibitem{Atiyah:1985fd}
M.~F. Atiyah and N.~J. Hitchin, {\it {Low-energy scattering of nonAbelian
  magnetic monopoles}},  {\em Phil. Trans. Roy. Soc. Lond. A} {\bf 315} (1985)
  459--469.

\bibitem{Sen:1997kz}
A.~Sen, {\it {A Note on enhanced gauge symmetries in M and string theory}},
  {\em JHEP} {\bf 09} (1997) 001,
  [\href{http://arxiv.org/abs/hep-th/9707123}{{\tt hep-th/9707123}}].

\bibitem{Cicoli:2022vny}
M.~Cicoli, M.~Licheri, R.~Mahanta, and A.~Maharana, {\it {Flux vacua with
  approximate flat directions}},  {\em JHEP} {\bf 10} (2022) 086,
  [\href{http://arxiv.org/abs/2209.02720}{{\tt arXiv:2209.02720}}].

\bibitem{Gao:2020xqh}
X.~Gao, A.~Hebecker, and D.~Junghans, {\it {Control issues of KKLT}},  {\em
  Fortsch. Phys.} {\bf 68} (2020) 2000089,
  [\href{http://arxiv.org/abs/2009.03914}{{\tt arXiv:2009.03914}}].

\bibitem{Carta:2021lqg}
F.~Carta and J.~Moritz, {\it {Resolving spacetime singularities in flux
  compactifications \& KKLT}},  {\em JHEP} {\bf 08} (2021) 093,
  [\href{http://arxiv.org/abs/2101.05281}{{\tt arXiv:2101.05281}}].

\bibitem{Honma:2021klo}
Y.~Honma and H.~Otsuka, {\it {Small flux superpotential in F-theory
  compactifications}},  {\em Phys. Rev. D} {\bf 103} (2021), no.~12 126022,
  [\href{http://arxiv.org/abs/2103.03003}{{\tt arXiv:2103.03003}}].

\bibitem{Marchesano:2021gyv}
F.~Marchesano, D.~Prieto, and M.~Wiesner, {\it {F-theory flux vacua at large
  complex structure}},  {\em JHEP} {\bf 08} (2021) 077,
  [\href{http://arxiv.org/abs/2105.09326}{{\tt arXiv:2105.09326}}].

\bibitem{Candelas:1990rm}
P.~Candelas, X.~C. De~La~Ossa, P.~S. Green, and L.~Parkes, {\it {A Pair of
  Calabi-Yau manifolds as an exactly soluble superconformal theory}},  {\em
  Nucl. Phys. B} {\bf 359} (1991) 21--74.

\bibitem{Candelas:1993dm}
P.~Candelas, X.~De~La~Ossa, A.~Font, S.~H. Katz, and D.~R. Morrison, {\it
  {Mirror symmetry for two parameter models. 1.}},  {\em Nucl. Phys. B} {\bf
  416} (1994) 481--538, [\href{http://arxiv.org/abs/hep-th/9308083}{{\tt
  hep-th/9308083}}].

\bibitem{Hosono:1993qy}
S.~Hosono, A.~Klemm, S.~Theisen, and S.-T. Yau, {\it {Mirror symmetry, mirror
  map and applications to Calabi-Yau hypersurfaces}},  {\em Commun. Math.
  Phys.} {\bf 167} (1995) 301--350,
  [\href{http://arxiv.org/abs/hep-th/9308122}{{\tt hep-th/9308122}}].

\bibitem{Candelas:1994hw}
P.~Candelas, A.~Font, S.~H. Katz, and D.~R. Morrison, {\it {Mirror symmetry for
  two parameter models. 2.}},  {\em Nucl. Phys. B} {\bf 429} (1994) 626--674,
  [\href{http://arxiv.org/abs/hep-th/9403187}{{\tt hep-th/9403187}}].

\bibitem{Hosono:1994ax}
S.~Hosono, A.~Klemm, S.~Theisen, and S.-T. Yau, {\it {Mirror symmetry, mirror
  map and applications to complete intersection Calabi-Yau spaces}},  {\em
  Nucl. Phys. B} {\bf 433} (1995) 501--554,
  [\href{http://arxiv.org/abs/hep-th/9406055}{{\tt hep-th/9406055}}].

\bibitem{Gopakumar:1998ii}
R.~Gopakumar and C.~Vafa, {\it {M theory and topological strings. 1.}},
  \href{http://arxiv.org/abs/hep-th/9809187}{{\tt hep-th/9809187}}.

\bibitem{Gopakumar:1998jq}
R.~Gopakumar and C.~Vafa, {\it {M theory and topological strings. 2.}},
  \href{http://arxiv.org/abs/hep-th/9812127}{{\tt hep-th/9812127}}.

\bibitem{Gukov:1999ya}
S.~Gukov, C.~Vafa, and E.~Witten, {\it {CFT's from Calabi-Yau four folds}},
  {\em Nucl. Phys. B} {\bf 584} (2000) 69--108,
  [\href{http://arxiv.org/abs/hep-th/9906070}{{\tt hep-th/9906070}}]. [Erratum:
  Nucl.Phys.B 608, 477--478 (2001)].

\bibitem{Giddings:2005ff}
S.~B. Giddings and A.~Maharana, {\it {Dynamics of warped compactifications and
  the shape of the warped landscape}},  {\em Phys. Rev. D} {\bf 73} (2006)
  126003, [\href{http://arxiv.org/abs/hep-th/0507158}{{\tt hep-th/0507158}}].

\bibitem{Witten:1996bn}
E.~Witten, {\it {Nonperturbative superpotentials in string theory}},  {\em
  Nucl. Phys. B} {\bf 474} (1996) 343--360,
  [\href{http://arxiv.org/abs/hep-th/9604030}{{\tt hep-th/9604030}}].

\bibitem{Frey:2002hf}
A.~R. Frey and J.~Polchinski, {\it {N=3 warped compactifications}},  {\em Phys.
  Rev. D} {\bf 65} (2002) 126009,
  [\href{http://arxiv.org/abs/hep-th/0201029}{{\tt hep-th/0201029}}].

\bibitem{Strominger:1996it}
A.~Strominger, S.-T. Yau, and E.~Zaslow, {\it {Mirror symmetry is T duality}},
  {\em Nucl. Phys. B} {\bf 479} (1996) 243--259,
  [\href{http://arxiv.org/abs/hep-th/9606040}{{\tt hep-th/9606040}}].

\bibitem{Gurrieri:2002wz}
S.~Gurrieri, J.~Louis, A.~Micu, and D.~Waldram, {\it {Mirror symmetry in
  generalized Calabi-Yau compactifications}},  {\em Nucl. Phys. B} {\bf 654}
  (2003) 61--113, [\href{http://arxiv.org/abs/hep-th/0211102}{{\tt
  hep-th/0211102}}].

\bibitem{Caviezel:2008ik}
C.~Caviezel, P.~Koerber, S.~Kors, D.~Lust, D.~Tsimpis, and M.~Zagermann, {\it
  {The Effective theory of type IIA AdS(4) compactifications on nilmanifolds
  and cosets}},  {\em Class. Quant. Grav.} {\bf 26} (2009) 025014,
  [\href{http://arxiv.org/abs/0806.3458}{{\tt arXiv:0806.3458}}].

\bibitem{Kaste:2003dh}
P.~Kaste, R.~Minasian, M.~Petrini, and A.~Tomasiello, {\it {Nontrivial RR two
  form field strength and SU(3) structure}},  {\em Fortsch. Phys.} {\bf 51}
  (2003) 764--768, [\href{http://arxiv.org/abs/hep-th/0301063}{{\tt
  hep-th/0301063}}].

\bibitem{Behrndt:2004mj}
K.~Behrndt and M.~Cvetic, {\it {General N=1 supersymmetric fluxes in massive
  type IIA string theory}},  {\em Nucl. Phys. B} {\bf 708} (2005) 45--71,
  [\href{http://arxiv.org/abs/hep-th/0407263}{{\tt hep-th/0407263}}].

\bibitem{Grana:2005jc}
M.~Grana, {\it {Flux compactifications in string theory: A Comprehensive
  review}},  {\em Phys. Rept.} {\bf 423} (2006) 91--158,
  [\href{http://arxiv.org/abs/hep-th/0509003}{{\tt hep-th/0509003}}].

\bibitem{Marchesano:2006ns}
F.~Marchesano, {\it {D6-branes and torsion}},  {\em JHEP} {\bf 05} (2006) 019,
  [\href{http://arxiv.org/abs/hep-th/0603210}{{\tt hep-th/0603210}}].

\bibitem{Sorkin:1983ns}
R.~d. Sorkin, {\it {Kaluza-Klein Monopole}},  {\em Phys. Rev. Lett.} {\bf 51}
  (1983) 87--90.

\bibitem{Gross:1983hb}
D.~J. Gross and M.~J. Perry, {\it {Magnetic Monopoles in Kaluza-Klein
  Theories}},  {\em Nucl. Phys. B} {\bf 226} (1983) 29--48.

\bibitem{Seiberg:1996nz}
N.~Seiberg and E.~Witten, {\it {Gauge dynamics and compactification to
  three-dimensions}},  in {\em {Conference on the Mathematical Beauty of
  Physics (In Memory of C. Itzykson)}}, pp.~333--366, 6, 1996.
\newblock \href{http://arxiv.org/abs/hep-th/9607163}{{\tt hep-th/9607163}}.

\bibitem{Witten:1995ex}
E.~Witten, {\it {String theory dynamics in various dimensions}},  {\em Nucl.
  Phys. B} {\bf 443} (1995) 85--126,
  [\href{http://arxiv.org/abs/hep-th/9503124}{{\tt hep-th/9503124}}].

\bibitem{Kaste:2003zd}
P.~Kaste, R.~Minasian, and A.~Tomasiello, {\it {Supersymmetric M theory
  compactifications with fluxes on seven-manifolds and G structures}},  {\em
  JHEP} {\bf 07} (2003) 004, [\href{http://arxiv.org/abs/hep-th/0303127}{{\tt
  hep-th/0303127}}].

\bibitem{Behrndt:2005im}
K.~Behrndt, M.~Cvetic, and T.~Liu, {\it {Classification of supersymmetric flux
  vacua in M theory}},  {\em Nucl. Phys. B} {\bf 749} (2006) 25--68,
  [\href{http://arxiv.org/abs/hep-th/0512032}{{\tt hep-th/0512032}}].

\bibitem{Ruback:1986ag}
P.~J. Ruback, {\it {The Motion of {Kaluza-Klein} Monopoles}},  {\em Commun.
  Math. Phys.} {\bf 107} (1986) 93--102.

\bibitem{Hanany:2000fw}
A.~Hanany and B.~Pioline, {\it {(Anti-)instantons and the Atiyah-Hitchin
  manifold}},  {\em JHEP} {\bf 07} (2000) 001,
  [\href{http://arxiv.org/abs/hep-th/0005160}{{\tt hep-th/0005160}}].

\bibitem{Frey:2013bha}
A.~R. Frey and J.~Roberts, {\it {The Dimensional Reduction and K{\"a}hler
  Metric of Forms In Flux and Warping}},  {\em JHEP} {\bf 10} (2013) 021,
  [\href{http://arxiv.org/abs/1308.0323}{{\tt arXiv:1308.0323}}].

\bibitem{Candelas:2016fdy}
P.~Candelas, A.~Constantin, and C.~Mishra, {\it {Calabi-Yau Threefolds with
  Small Hodge Numbers}},  {\em Fortsch. Phys.} {\bf 66} (2018), no.~6 1800029,
  [\href{http://arxiv.org/abs/1602.06303}{{\tt arXiv:1602.06303}}].

\bibitem{Gukov:2002jv}
S.~Gukov, S.-T. Yau, and E.~Zaslow, {\it {Duality and fibrations on G(2)
  manifolds}},  \href{http://arxiv.org/abs/hep-th/0203217}{{\tt
  hep-th/0203217}}.

\bibitem{Grana:2005sn}
M.~Grana, R.~Minasian, M.~Petrini, and A.~Tomasiello, {\it {Generalized
  structures of N=1 vacua}},  {\em JHEP} {\bf 11} (2005) 020,
  [\href{http://arxiv.org/abs/hep-th/0505212}{{\tt hep-th/0505212}}].

\bibitem{Grana:2000jj}
M.~Grana and J.~Polchinski, {\it {Supersymmetric three form flux perturbations
  on AdS(5)}},  {\em Phys. Rev. D} {\bf 63} (2001) 026001,
  [\href{http://arxiv.org/abs/hep-th/0009211}{{\tt hep-th/0009211}}].

\bibitem{Green:1996dd}
M.~B. Green, J.~A. Harvey, and G.~W. Moore, {\it {I-brane inflow and anomalous
  couplings on d-branes}},  {\em Class. Quant. Grav.} {\bf 14} (1997) 47--52,
  [\href{http://arxiv.org/abs/hep-th/9605033}{{\tt hep-th/9605033}}].

\bibitem{Cheung:1997az}
Y.-K.~E. Cheung and Z.~Yin, {\it {Anomalies, branes, and currents}},  {\em
  Nucl. Phys. B} {\bf 517} (1998) 69--91,
  [\href{http://arxiv.org/abs/hep-th/9710206}{{\tt hep-th/9710206}}].

\bibitem{Minasian:1997mm}
R.~Minasian and G.~W. Moore, {\it {K theory and Ramond-Ramond charge}},  {\em
  JHEP} {\bf 11} (1997) 002, [\href{http://arxiv.org/abs/hep-th/9710230}{{\tt
  hep-th/9710230}}].

\bibitem{Grimm:2004ua}
T.~W. Grimm and J.~Louis, {\it {The Effective action of type IIA Calabi-Yau
  orientifolds}},  {\em Nucl. Phys. B} {\bf 718} (2005) 153--202,
  [\href{http://arxiv.org/abs/hep-th/0412277}{{\tt hep-th/0412277}}].

\bibitem{Brunner:2003zm}
I.~Brunner and K.~Hori, {\it {Orientifolds and mirror symmetry}},  {\em JHEP}
  {\bf 11} (2004) 005, [\href{http://arxiv.org/abs/hep-th/0303135}{{\tt
  hep-th/0303135}}].

\bibitem{Acharya:2001gy}
B.~S. Acharya and E.~Witten, {\it {Chiral fermions from manifolds of G(2)
  holonomy}},  \href{http://arxiv.org/abs/hep-th/0109152}{{\tt
  hep-th/0109152}}.

\bibitem{Pantev:2009de}
T.~Pantev and M.~Wijnholt, {\it {Hitchin's Equations and M-Theory
  Phenomenology}},  {\em J. Geom. Phys.} {\bf 61} (2011) 1223--1247,
  [\href{http://arxiv.org/abs/0905.1968}{{\tt arXiv:0905.1968}}].

\bibitem{Harvey:1999as}
J.~A. Harvey and G.~W. Moore, {\it {Superpotentials and membrane instantons}},
  \href{http://arxiv.org/abs/hep-th/9907026}{{\tt hep-th/9907026}}.

\bibitem{Kim:2022jvv}
M.~Kim, {\it {D-instanton superpotential in string theory}},  {\em JHEP} {\bf
  03} (2022) 054, [\href{http://arxiv.org/abs/2201.04634}{{\tt
  arXiv:2201.04634}}].

\bibitem{Alexandrov:2022mmy}
S.~Alexandrov, A.~H. F\i{}rat, M.~Kim, A.~Sen, and B.~Stefa\'nski, {\it
  {D-instanton induced superpotential}},  {\em JHEP} {\bf 07} (2022) 090,
  [\href{http://arxiv.org/abs/2204.02981}{{\tt arXiv:2204.02981}}].

\bibitem{Kim:2022uni}
M.~Kim, {\it {On D3-brane Superpotential}},
  \href{http://arxiv.org/abs/2207.01440}{{\tt arXiv:2207.01440}}.

\bibitem{Jefferson:2022ssj}
P.~Jefferson and M.~Kim, {\it {On the intermediate Jacobian of M5-branes}},
  {\em JHEP} {\bf 05} (2024) 180, [\href{http://arxiv.org/abs/2211.00210}{{\tt
  arXiv:2211.00210}}].

\bibitem{Kim:2023cbh}
M.~Kim, {\it {D-instanton, threshold corrections, and topological string}},
  {\em JHEP} {\bf 05} (2023) 097, [\href{http://arxiv.org/abs/2301.03602}{{\tt
  arXiv:2301.03602}}].

\bibitem{Shiu:2008ry}
G.~Shiu, G.~Torroba, B.~Underwood, and M.~R. Douglas, {\it {Dynamics of Warped
  Flux Compactifications}},  {\em JHEP} {\bf 06} (2008) 024,
  [\href{http://arxiv.org/abs/0803.3068}{{\tt arXiv:0803.3068}}].

\bibitem{Frey:2008xw}
A.~R. Frey, G.~Torroba, B.~Underwood, and M.~R. Douglas, {\it {The Universal
  Kahler Modulus in Warped Compactifications}},  {\em JHEP} {\bf 01} (2009)
  036, [\href{http://arxiv.org/abs/0810.5768}{{\tt arXiv:0810.5768}}].

\bibitem{Martucci:2016pzt}
L.~Martucci, {\it {Warped K\"ahler potentials and fluxes}},  {\em JHEP} {\bf
  01} (2017) 056, [\href{http://arxiv.org/abs/1610.02403}{{\tt
  arXiv:1610.02403}}].

\bibitem{Junghans:2023lpo}
D.~Junghans, {\it {de Sitter-eating O-planes in supercritical string theory}},
  {\em JHEP} {\bf 12} (2023) 196, [\href{http://arxiv.org/abs/2308.00026}{{\tt
  arXiv:2308.00026}}].

\bibitem{Anderson:2020hux}
L.~B. Anderson, M.~Gerdes, J.~Gray, S.~Krippendorf, N.~Raghuram, and F.~Ruehle,
  {\it {Moduli-dependent Calabi-Yau and SU(3)-structure metrics from Machine
  Learning}},  {\em JHEP} {\bf 05} (2021) 013,
  [\href{http://arxiv.org/abs/2012.04656}{{\tt arXiv:2012.04656}}].

\bibitem{Douglas:2020hpv}
M.~R. Douglas, S.~Lakshminarasimhan, and Y.~Qi, {\it {Numerical Calabi-Yau
  metrics from holomorphic networks}},
  \href{http://arxiv.org/abs/2012.04797}{{\tt arXiv:2012.04797}}.

\bibitem{Jejjala:2020wcc}
V.~Jejjala, D.~K. Mayorga~Pena, and C.~Mishra, {\it {Neural network
  approximations for Calabi-Yau metrics}},  {\em JHEP} {\bf 08} (2022) 105,
  [\href{http://arxiv.org/abs/2012.15821}{{\tt arXiv:2012.15821}}].

\bibitem{Larfors:2022nep}
M.~Larfors, A.~Lukas, F.~Ruehle, and R.~Schneider, {\it {Numerical metrics for
  complete intersection and Kreuzer\textendash{}Skarke Calabi\textendash{}Yau
  manifolds}},  {\em Mach. Learn. Sci. Tech.} {\bf 3} (2022), no.~3 035014,
  [\href{http://arxiv.org/abs/2205.13408}{{\tt arXiv:2205.13408}}].

\bibitem{Gerdes:2022nzr}
M.~Gerdes and S.~Krippendorf, {\it {CYJAX: A package for Calabi-Yau metrics
  with JAX}},  {\em Mach. Learn. Sci. Tech.} {\bf 4} (2023), no.~2 025031,
  [\href{http://arxiv.org/abs/2211.12520}{{\tt arXiv:2211.12520}}].

\bibitem{Halverson:2023ndu}
J.~Halverson and F.~Ruehle, {\it {Metric flows with neural networks}},  {\em
  Mach. Learn. Sci. Tech.} {\bf 5} (2024), no.~4 045020,
  [\href{http://arxiv.org/abs/2310.19870}{{\tt arXiv:2310.19870}}].

\bibitem{Lukas:2003dn}
A.~Lukas and S.~Morris, {\it {Moduli Kahler potential for M theory on a G(2)
  manifold}},  {\em Phys. Rev. D} {\bf 69} (2004) 066003,
  [\href{http://arxiv.org/abs/hep-th/0305078}{{\tt hep-th/0305078}}].

\bibitem{Kachru:2001je}
S.~Kachru and J.~McGreevy, {\it {M theory on manifolds of G(2) holonomy and
  type IIA orientifolds}},  {\em JHEP} {\bf 06} (2001) 027,
  [\href{http://arxiv.org/abs/hep-th/0103223}{{\tt hep-th/0103223}}].

\bibitem{Anderson:2016cdu}
L.~B. Anderson, X.~Gao, J.~Gray, and S.-J. Lee, {\it {Multiple Fibrations in
  Calabi-Yau Geometry and String Dualities}},  {\em JHEP} {\bf 10} (2016) 105,
  [\href{http://arxiv.org/abs/1608.07555}{{\tt arXiv:1608.07555}}].

\bibitem{Kachru:2002gs}
S.~Kachru, J.~Pearson, and H.~L. Verlinde, {\it {Brane / flux annihilation and
  the string dual of a nonsupersymmetric field theory}},  {\em JHEP} {\bf 06}
  (2002) 021, [\href{http://arxiv.org/abs/hep-th/0112197}{{\tt
  hep-th/0112197}}].

\bibitem{Atiyah:2000zz}
M.~Atiyah, J.~M. Maldacena, and C.~Vafa, {\it {An M theory flop as a large N
  duality}},  {\em J. Math. Phys.} {\bf 42} (2001) 3209--3220,
  [\href{http://arxiv.org/abs/hep-th/0011256}{{\tt hep-th/0011256}}].

\bibitem{Becker:2004qh}
M.~Becker, K.~Dasgupta, A.~Knauf, and R.~Tatar, {\it {Geometric transitions,
  flops and nonKahler manifolds. I.}},  {\em Nucl. Phys. B} {\bf 702} (2004)
  207--268, [\href{http://arxiv.org/abs/hep-th/0403288}{{\tt hep-th/0403288}}].

\bibitem{Becker:2005ef}
M.~Becker, K.~Dasgupta, S.~H. Katz, A.~Knauf, and R.~Tatar, {\it {Geometric
  transitions, flops and non-Kahler manifolds. II.}},  {\em Nucl. Phys. B} {\bf
  738} (2006) 124--183, [\href{http://arxiv.org/abs/hep-th/0511099}{{\tt
  hep-th/0511099}}].

\end{thebibliography}\endgroup
\bibliographystyle{JHEP}
\end{document}